\documentclass[twocolumn,preprintnumbers]{revtex4}

\usepackage{graphicx}
\usepackage{epsf}
\usepackage{amsmath,amssymb}
\usepackage{color}
\usepackage{ulem}
\newcommand{\bvec}{\boldsymbol}

\begin{document}

\title{Short-range and tensor correlations in $^4$He and $^8$Be
studied with antisymmetrized quasi cluster model
}

\author{N. Itagaki}

\affiliation{
Yukawa Institute for Theoretical Physics, Kyoto University,
Kitashirakawa Oiwake-Cho, Kyoto 606-8502, Japan
}

\author{H. Matsuno, Y. Kanada-En'yo}

\affiliation{
Department of Physics, Kyoto University, Kitashirakawa Oiwake-Cho, Kyoto 606-8502, Japan
}

\begin{abstract}
We apply tensor version of antisymmetrized quasi cluster model (AQCM-T) to
$^4\textrm{He}$ and $^8\textrm{Be}$
while focusing on the
$NN$ correlations in $\alpha$ clusters. 
We adopt the $NN$ interactions including
realistic ones containing a repulsive core for the central part in addition to the tensor part.
In $^4\textrm{He}$,
the $pn$ pair in the $^3D$ channel has been known to play a decisive role in
the tensor correlation and the framework is capable of treating not only this channel but also the $NN$ 
correlations in the $^1S$ and $^3S$ channels.
In $^8\textrm{Be}$,
when two $\alpha$ clusters approach, the $^3D$ pair is suppressed
because of the Pauli blocking effect, which 
induces the decrease of the $^3S$ component through the $^3S$-$^3D$ coupling.
This effect results in the reduction of the attractive effect of the central-even interaction
in the middle-range region.

\end{abstract}
\maketitle

\section{Introduction} \label{sec:introduction}

The $^4$He nucleus is a strongly bound many-nucleon system
in the light mass region, thus the $\alpha$ clusters
can be basic building blocks of the nuclear structure.
The $\alpha$ cluster models ~\cite{Brink,Fujiwara} have been developed and applied 
in numerous works
for the description of 
cluster structures such as $3\alpha$ clustering in
the so-called Hoyle state of $^{12}$C~\cite{Hoyle,Uegaki,THSR}.
In most of the conventional
cluster models, however, each $\alpha$ cluster is assumed as 
a simple $(0s)^4$ configuration, 
which is spin singlet,
and therefore the contributions of non-central interactions, the spin-orbit and tensor interactions, 
completely vanish
even though they play crucial roles 
in the nuclear structure~\cite{ATMS,Kamada,TOSM}.
Also, nucleons are correlated owing to the repulsive core in the short-range part of the central interaction,
and this effect is not explicitly treated in the conventional structure models including cluster models. 
These days, such $NN$ correlation is widely discussed 
based on modern {\it ab initio} theories not only in very light nuclei
but also in medium-heavy nuclei~\cite{Alvioli,Feldmeier,Wiringa,Atti,Alvioli-2}. 
Moreover, there are many experimental attempts  with deep inelastic scattering 
to pin down the $NN$ correlations in 
nuclei in wide mass number regions (see for example, Refs.~\cite{Atti,Hen:2016kwk} and references therein).
In study of cluster aspects in nuclear systems, 
it is an urgent  issue  to extend the model space and take into account these higher correlations beyond the $(0s)^4$ configuration
in the cluster model.

Concerning the non-central interactions, recently, 
many attempts of directly taking into account them
for the microscopic studies of cluster structure have begun; 
antisymmetrized molecular dynamics (AMD)~\cite{KanadaEnyo:1995tb,KanadaEnyo:1995ir,AMDsupp,KanadaEn'yo:2012bj},
its extended version~\cite{Dote}, and
Fermionic molecular dynamics (FMD)~\cite{Neff,Roth,Chernykh} combined with the 
unitary correlation method (UCOM).
In UCOM, 
the effects of non-central interaction  and also short-range correlation are
included with the unitary transformation of the Hamiltonian,
which 
in principle induces many-body operators up to $A$ (mass number) body.

Our aim is to
introduce an effective model, which is phenomenological but capable of
directly taking into account the non-central interactions in a simplified manner
without transforming the Hamiltonian. 
Concerning the rank one non-central interaction,
the spin-orbit interaction,  
we proposed the antisymmetrized quasi cluster model 
(AQCM)~\cite{Simple,Itagaki-SMT,Masui,Yoshida2,Ne-Mg,Suhara,Suhara2015,Itagaki,Itagaki-CO,Matsuno,Matsuno2,O24}.
By introducing a parameter for the imaginary part of the Gaussian centroids of 
single-particle wave functions in  $\alpha$ clusters,
we can smoothly transform $\alpha$ clusters to
$jj$-coupling shell model wave functions, where
the transformed $\alpha$ clusters contains cluster breaking components and 
are called quasi clusters.

Recently, the imaginary centered Gaussian wave packets have been utilized to 
directly take into account the rank two non-central interaction, 
tensor interaction~\cite{iSMT,hmAMD,hmAMD-2}.
In the previous paper~\cite{AQCM-T}, we newly proposed AQCM-T, 
which is an improved version of AQCM  so as to explicitly treat the tensor correlation
in the two-nucleon pairs.
It has been known that the $pn$ pair in the $^3D$ channel plays a decisive role in
the tensor correlation and this new framework is capable of treating not only this channel but also the $NN$ 
correlations in the $^1S$ and $^3S$ channels in a very simplified way.
The AQCM-T model with the new interaction was applied to $^8\textrm{Be}$,
where the relation between the $\alpha$-$\alpha$ cluster structure
and the tensor interaction has been discussed.
It was found that the tensor suppression gives a significant contribution 
to the $\alpha$-$\alpha$ repulsion at short distances consistently with the pioneering works
~\cite{Bando,QMC,Yamamoto}.

In the present study, we aim
to describe the short-range correlation 
caused by the repulsive core of the central interaction in addition to the tensor correlation;
we apply AQCM-T to
$^4\textrm{He}$ and $^8\textrm{Be}$ using
a realistic $NN$ interaction containing a repulsive core for the central part.
We focus on the
$NN$ correlations in $\alpha$ clusters.
This paper is organized as follows;
in Sec.~{\bf II}, the framework,
especially for the model wave function, is explained.
In Sec.~{\bf III}, the Hamiltonian of the present model is described.
In sections {\bf IV} and {\bf V}, the numerical results for $^4\textrm{He}$
and $^4\textrm{Be}$ are presented, respectively. 
The summary is presented in Sec.~{\bf VI}.

\section{Formulations}  \label{sec:formulation}

In the present study, we aim
to describe the short-range correlation 
caused by the repulsive core of the central interaction in addition to the tensor correlation;
we apply AQCM-T to
$^4\textrm{He}$ and $^8\textrm{Be}$ using
a realistic $NN$ interaction containing a repulsive core for the central part.
The procedures of the calculation based on AQCM-T  
is, in principle, the same as those presented in the previous paper~\cite{AQCM-T},
and the readers can refer to it
for the detailed formulations.
Here we first explain the AQCM-T treatment for a single $NN$-pair,
which are correlated, and next give the formulation of AQCM for actual nuclei,
$^4\textrm{He}$ and $^8\textrm{Be}$.

\subsection{AQCM-T for a $NN$-pair}

Each single-particle wave function is
written by Gaussian wave packet as 
\begin{eqnarray}
&&\psi_j(i)=\phi_{\bvec{S}_j}(\bvec{r}_i)\chi_j(s_i,\tau_i),\\
&&\phi_{\bvec{S}_j}(\bvec{r}_i)
=\left(\frac{2\nu}{\pi}\right)^{\frac{3}{4}}
e^{-\nu(\bvec{r}_i-\bvec{S}_{j})^2},
\label{spwf}
\end{eqnarray}
where 
$\bvec{S}_j$ ($j=1,2$)
is the Gaussian centroid and
$\chi_j$ is the spin-isospin wave function.
The width parameter $\nu$ is set to $\nu=0.25$ fm$^{-2}$ 
and fixed in all calculations of this paper. 

For Gaussian centroids of two nucleons in a correlated $NN$-pair, 
we adopt the following complex conjugate values, 
\begin{eqnarray}
\bvec{S}_{1}=\bvec{R}+\frac{i\bvec{K}}{\nu}
=\bvec{R}+\frac{i\bvec{k}}{2\nu}, \nonumber \\
\bvec{S}_{2}=\bvec{R}-\frac{i\bvec{K}}{\nu}
=\bvec{R}-\frac{i\bvec{k}}{2\nu},
\end{eqnarray}
where $\bvec{R}$ and $\bvec{K}$ are real vectors, and $\bvec{k}\equiv2\bvec{K}$.
The $NN$-pair wave function can be written in a separated form for
the relative and center of mass (cm) 
coordinates, $\bvec{r}=\bvec{r}_1-\bvec{r}_2$ and $\bvec{r}_g=(\bvec{r}_1+\bvec{r}_2)/2$, respectively, 
as 
\begin{eqnarray}
&&\phi_{\bvec{S}_1}(\bvec{r}_1)\phi_{\bvec{S}_2} (\bvec{r}_2)= \varphi_{\bvec{k}}(\bvec{r})  \phi_g(\bvec{r}_g),  \\
&&\varphi_{\bvec{k}}(\bvec{r})=\left(\frac{\nu}{\pi}\right)^{\frac{3}{4}}
e^{-\frac{\nu}{2}r^2+i\bvec{k}\cdot\bvec{r}
+\frac{k^2}{2\nu}},  \label{eq:phi} \\
&& \phi_g(\bvec{r}_g)
=\left(\frac{4\nu}{\pi}\right)^{\frac{3}{4}}
e^{-2\nu(\bvec{r}_g-\bvec{R})^2}. 
\end{eqnarray}
The expectation values of the positions and momenta of the relative and cm coordinates 
are given as  
\begin{eqnarray}
&&\langle \widehat{\bvec{r}}\rangle =0, \qquad \langle \widehat{\bvec{p}}\rangle =\bvec{k},\\
&&\langle \widehat{\bvec{r}}_{g}\rangle =\bvec{R} , \qquad \langle \widehat{\bvec{p}}_{g}\rangle =0.
\end{eqnarray}
Note that the Fourier components of  the relative wave function $\varphi_{\bvec{k}}(\bvec{r})$
also has a Gaussian form, which is
localized at $\bvec{k}$ with the dispersion of $\sqrt{\nu}$.
As shown in Ref.~\cite{hmAMD}, 
the relative wave function for a non-zero vector $\bvec{k}$ contain various partial-wave components as 
\begin{eqnarray}
\varphi_{\bvec{k}}(\bvec{r}) &&=4\pi
\left(\frac{\nu}{\pi}\right)^{\frac{3}{4}}
e^{-\frac{\nu}{2}r^2+\frac{k^2}{2\nu}} \nonumber \\&&\times \sum_{lm} i^l j_l(kr) Y_{lm}(\bvec{e}_k)
 Y_{lm}(\bvec{e}_r).
\end{eqnarray}
 In order to take into account the tensor coupling between 
$^3D$ and $^3S$ channels of the $T=0$ $NN$-pair, 
we project the $NN$-pair onto the
positive-parity state and set $\bvec{k}$ along the $z$ axis as $\bvec{k}=(0,0,k)$
as done in the previous paper~\cite{AQCM-T}. Then
the relative wave function can be expressed as 
\begin{eqnarray}
&&\varphi^+_k(\bvec{r})=
\left(\frac{\nu}{\pi}\right)^{\frac{3}{4}}
e^{-\frac{\nu}{2}r^2+\frac{k^2}{2\nu}}\cos(kz)\nonumber\\
&&=\left(\frac{\nu}{\pi}\right)^{\frac{3}{4}}
 e^{-\frac{\nu}{2}r^2+\frac{k^2}{2\nu}}
4\pi\sum_{l=\textrm{even}} \sqrt{\frac{2l+1}{4\pi}} i^l j_l(kr) Y_{l0}(\bvec{e}_r)\nonumber\\
&&=\sum_{l=\textrm{even}} 
a_l \varphi^{(l)}_k(r) Y_{l0}(\bvec{e}_r),
\label{eq:partial-NN}
\end{eqnarray}
where $a_l$ is the normalization 
factor, and 
$\varphi^{(l)}_k(r)$ is the normalized radial wave function 
of the $l$-even basis state
proportional to $e^{-\frac{\nu}{2}r^2} j_l(kr)$.


\subsection{AQCM-T for $^4\textrm{He}$} 

Next, we apply the $NN$-pair wave function described in the previous subsection 
to the two correlating nucleons in $^4\textrm{He}$.

\subsubsection{Model wave function of $^4\textrm{He}$}

For the $^4\textrm{He}$ system, 
in addition to the correlated $NN$-pair described by AQCM-T, 
we consider a $(0s)^2$ (uncorrelated) pair, and both pairs are placed at the origin. 
The AQCM-T wave function for $^4\textrm{He}$ is expressed as 
\begin{eqnarray} \label{eq:4He}
\Phi^\textrm{AQCM-T}_{^4\textrm{He},0^+}&&=\widehat{P}^{0+}{\cal A}\{\phi_{\frac{i\bvec{k}}{2\nu}}\chi_1
\phi_{-\frac{i\bvec{k}}{2\nu}}\chi_2,\phi_{0}\chi_3,\phi_{0}\chi_4\} \nonumber \\
&&
=\widehat{P}^{0+}{\cal A}\{\phi_{\frac{i\bvec{k}}{2\nu}}
\phi_{-\frac{i\bvec{k}}{2\nu}}\phi_{0}\phi_{0}\otimes \chi_1\chi_2\chi_3\chi_4\}, \nonumber\\
\end{eqnarray}
where ${\cal A}$ is the antisymmetrizer for all the nucleons, $\widehat{P}^{0+}$ is the 
projection operator to
$J^\pi=0^+$ (in practice numerically performed), 
and $\phi_{0}=\phi_{\bvec{S}=0}$ is the 
spatial wave function for a nucleon in the
$0s$ orbit.
The spatial wave function of the total system in the intrinsic frame before the projections
is rewritten as
\begin{eqnarray}
&&\phi_{\frac{i\bvec{k}}{2\nu}}
\phi_{-\frac{i\bvec{k}}{2\nu}}\phi_{0}\phi_{0}
=
\phi_g(\bvec{r}_{g})\phi_g(\bvec{r}'_{g})
\varphi_{\bvec{k}}(\bvec{r})\varphi_{0}(\bvec{r}'),\label{eq:spatial} \\
&&\bvec{r}_g=\frac{\bvec{r}_1+\bvec{r}_2}{2},\qquad \bvec{r}'_g=\frac{\bvec{r}_3+\bvec{r}_4}{2},\\
&&\bvec{r}=\bvec{r}_1-\bvec{r}_2,\qquad \bvec{r}'=\bvec{r}_3-\bvec{r}_4,
\end{eqnarray}
where $\bvec{r}$ and $\bvec{r}_g$ ($\bvec{r}'$ and $\bvec{r}'_g$) are the relative and cm coordinates of the
pair, respectively.
The $NN$ correlation is taken into account through $\varphi_{\bvec{k}}(\bvec{r})$
in the case of the correlated $NN$-pair.

For the choice of the spin-isospin configurations,  
one should care about the redundancies originating
from the parity and angular momentum projections as well as the Fermi statistics 
(antisymmetrization effect). 
For the $J^\pi=0^+$ states of $^4\textrm{He}$,
the model space for a given $k $ value $(k \ne 0)$ contains 
only the  $S$-wave ($\varphi^{(0)}_k$) and 
$D$-wave ($\varphi^{(2)}_k$) components, which are
coupled to the total intrinsic spin $S=0$ and $S=2$ of 
the four nucleons, respectively. For the $0^+$ state
with the $\bvec{k}=(0,0,k)$ choice, 
only the $S_z=0$ states contribute,
and in total we have
five independent spin and isospin configurations; 
\begin{eqnarray} \label{eq:config5}
&&\chi_1 \chi_2 \chi_3\chi_4=\nonumber \\
&&\{  
{p}\uparrow {p}\downarrow n\uparrow n\downarrow, \ \
{n}\uparrow {n}\downarrow p\uparrow p\downarrow, \ \ \nonumber \\
&& 
p\uparrow n\uparrow p\downarrow n\downarrow, \ \
p\uparrow n\downarrow p\uparrow n\downarrow, \ \ \nonumber \\
&& p\uparrow n\downarrow p\downarrow n\uparrow
\}.
\end{eqnarray}

Furthermore, 
when we ignore small breaking of the isospin symmetry by the Coulomb interaction, the five configurations in Eq.~\eqref{eq:config5} can be
reduced into three channels with respect to spin and isospin symmetries of 
the $NN$-pair as 
\begin{eqnarray}  
\label{eq:channel3-1}
^1S:&& \phi_g(\bvec{r}_{g})\phi_g(\bvec{r}'_{g})\otimes
\varphi^{(0)}_k(r)\varphi^{(0)}_{0}(r')   \nonumber\\
&&
\otimes Y_{00}(\bvec{e}_r) Y_{00}(\bvec{e}_{r'})  \otimes \chi^\sigma_{0} \chi^\sigma_{0}
\otimes [ \chi^\tau_{1} \chi^\tau_{1}]_{T=0},
\end{eqnarray}
\begin{eqnarray}  
\label{eq:channel3-2}
^3S: && \phi_g(\bvec{r}_{g})\phi_g(\bvec{r}'_{g})\otimes
\varphi^{(0)}_k(r)\varphi^{(0)}_{0}(r')   \nonumber\\
&&\otimes
Y_{00}(\bvec{e}_r) Y_{00}(\bvec{e}_{r'})  \otimes [\chi^\sigma_{1} \chi^\sigma_{1}]_{S=0}\otimes
\chi^\tau_{0} \chi^\tau_{0},
\end{eqnarray}
\begin{eqnarray}  
\label{eq:channel3-3}
^3D:&& \phi_g(\bvec{r}_{g})\phi_g(\bvec{r}'_{g})\otimes
\varphi^{(2)}_k(r)\varphi^{(0)}_{0}(r')   \nonumber\\
&&\otimes\left[ Y_{20}(\bvec{e}_r) Y_{00}(\bvec{e}_{r'})  \otimes 
 [\chi^\sigma_{1} \chi^\sigma_{1}]_{S=2}\right]_{J=0}\otimes
\chi^\tau_{0} \chi^\tau_{0},\nonumber\\
\end{eqnarray}
where $\chi^\sigma_{0,1}$ ($\chi^\tau_{1,0}$) is the spin (isospin) function of the $NN$-pairs coupled to 
the spin (isospin) singlet and triplet states, respectively. 
Note that $\varphi^{(0)}_{0}(=\varphi^{(0)}_{k=0})$ expresses
the uncorrelated $NN$-pair with the $(0s)^2$ configuration.
The first (second) configuration
takes into account the $NN$ correlation in the $^1S$ ($^3S$) channel
and is essential for the short-range correlation caused by the repulsive core of the central interaction. 
The third configuration is the so-called $D$-state component and contributes to the
tensor correlation.
We call the first, second, and third configurations, the 
$^1S$, $^3S$, and $^3D$ channels, respectively.  

If the charge symmetry breaking by the Coulomb interaction 
can be ignored, 
the three-channel ($^1S$, $^3S$, and $^3D$) calculation 
is equivalent to the configuration mixing of the 
five configurations defined in Eq.~\eqref{eq:config5}. 
Indeed,  in the practical calculation of $^4\textrm{He}$, we found that
almost equivalent results are 
obtained in two cases indicating that the
charge symmetry breaking is negligibly small.

In the present framework,
$\Phi^{\textrm{AQCM-T}}_{^4\textrm{He},0^+}$ defined in Eqs.~\eqref{eq:4He} and \eqref{eq:spatial}
is 
a basis wave function
specified by the momentum parameter $k$ in $\bvec{k}=(0,0,k)$ 
and the channel $\beta=\{^1S$, $^3S$, and $^3D$\}.
The total wave function  for the ground state,
$\Psi_{^4\textrm{He},\textrm{g.s.}}$, 
is therefore expressed by linear combination of 
various $k$ values and the spin and isospin configurations as 
\begin{eqnarray}
\label{eq:gcm-4He}
\Psi_{^4\textrm{He},\textrm{g.s.}}=c_0
\Phi^{0s}_{^4\textrm{He}}+
 \sum_{k} \sum_\beta c(k,\beta) \Phi^{\textrm{AQCM-T}}_{^4\textrm{He},0^+}(k,\beta).
\end{eqnarray} 
Here, $\Phi^{0s}_{^4\textrm{He}}$ in the first term 
is the $(0s)^4$ wave function, which is equivalent to the {
$k=0$ AQCM-T wave function with  
$\beta={}^1S$ or $\beta={}^3S$.
The coefficients $c_0$ and $c(k,\beta)$ are determined by diagonalizing the 
norm and Hamiltonian matrices comprised of the basis wave functions.
The superposition of $k$ 
in Eq.~\eqref{eq:gcm-4He} is nothing but 
the expansion of the relative wave function of the correlated $NN$-pair 
in the momentum space with localized Gaussians at the mean momentum $\bvec{k}$, and that of $\beta=\{^1S$, $^3S$, and $^3D\}$ corresponds to the 
coupled-channel calculation of the three channels. 

We explicitly treat the
$NN$ correlations of only a single $NN$-pair among the four nucleons
but omit higher-order correlations, where more than two
nucleons are involved. 
This ansatz is supported by the four-body calculation of $^4\textrm{He}$ 
by Horii 
{\it et al.} in Ref.~\cite{Horii} showing that the  
$^3S$-$^3D$ coupling in a single $NN$-pair with $T=0$ is essential to
describe the tensor correlation in the ground state. 
However, this ansatz may not hold 
in the case of extremely hard 
core at the short-range region of the central interaction, 
which may promote the non-negligible amount of
higher-order effects.
To include higher-order correlations, 
recently, Myo {\it et al.} has proposed 
further improved framework 
with finite imaginary parts
for the Gaussian centroids~\cite{hmAMD-2}.
However,  one of the advantages of the present AQCM-T model
is that one can  analyze the contribution
and pair wave function in each of the three 
($^1S$, $^3S$, and $^3D$) channels because  
the ${}^4\textrm{He}$ wave function is explicitly expressed 
based on the spin-isospin symmetry of the two $NN$-pairs
in the present model.

\subsubsection{Parameter settings for $^4\textrm{He}$}
For the ground state of $^4\textrm{He}$
($^4\textrm{He}(\textrm{g.s.})$), 
we perform calculations with the three channels 
($\beta=\{^1S,^3S, ^3D\}$).
For each channel,  
the basis states with
$k=0.5,1.0,\ldots, 5.5$~fm$^{-1}$ 
(11 states) are adopted
in addition to the $(0s)^4$ configuration.
We also perform truncated calculations by
selecting parameters $k$ and/or channels $\beta$
to clarify the roles of high momentum components 
in each $^1S$, $^3S$, and $^3D$ channel. 

\subsection{AQCM-T for $^{8}\textrm{Be}$} \label{subsec:be8}
\subsubsection{Basis wave function for two-$\alpha$ system}
Our aim is to investigate effects of the $NN$ correlations in heavier nuclei.
Here we extend the AQCM-T framework to $^8\textrm{Be}$ 
with a two-$\alpha$ cluster structure, 
in which 
one of $\alpha$ clusters is changed from the $(0s)^4$ configuration to
the correlated $^4\textrm{He}$ wave function
previously explained.
We  label the correlated $\alpha$ cluster as $\alpha_k$, and another $\alpha$ cluster 
with
the $(0s)^4$ configuration is labeled as $\alpha_0$.
We place $\alpha_k$ at $\bvec{R}=\frac{\bvec{d}}{2}$ and 
$\alpha_0$ at $\bvec{R}'=-\frac{\bvec{d}}{2}$ 
with the relative distance of $d \equiv | \bvec{d} |$.
After the antisymmetrization, 
the two-$\alpha$ wave function projected to $0^+$ is
\begin{equation} 
\Phi^{\textrm{AQCM-T}}_{2\alpha,0^+}(k,\beta, \bvec{d})
=\widehat{P}^{0+}{\cal A}\left\{ 
\Phi_{\alpha_k}(k,\beta,\bvec{R}) 
\Phi_{\alpha_0}(\bvec{R}') 
\right\},
\label{eq:8Be}
\end{equation}
where $k$ and $\beta$ are the momentum parameter and 
the channel of the 
$\alpha_k$ cluster.  
The $\alpha_k$ and $\alpha_0$ clusters are expressed using 
the AQCM-T wave function for $^4$He as
\begin{eqnarray}
&&\Phi_{\alpha_k}(k,\beta,\bvec{R})=
\Phi^\textrm{AQCM-T}_{^4\textrm{He},+}(k,\beta,\bvec{R})\nonumber \\
&&=\frac{1+\widehat{P}_k}{2}{\cal A}\{\phi_{\bvec{R}+\frac{i\bvec{k}}{2\nu}}
\phi_{\bvec{R}-\frac{i\bvec{k}}{2\nu}}
\phi_{\bvec{R}}\phi_{\bvec{R}}
\otimes \chi_1\chi_2\chi_3\chi_4 \}, \\
&& \Phi_{\alpha_0}(\bvec{R}') = \Phi^{0s}_{^4\textrm{He}} (\bvec{R}') \nonumber \\
&&={\cal A}\{\phi_{\bvec{R}'}\phi_{\bvec{R}'}\phi_{\bvec{R}'}\phi_{\bvec{R}'}
\otimes {p\uparrow}{p\downarrow} {n\uparrow} {n\downarrow} \}.
\end{eqnarray}
Here $\bvec{k}=(0,0, k)$,
and the operator $\widehat{P}_k$ transforms the imaginary part of
the correlated $NN$-pair as $k\to -k$, and 
$(1+\widehat{P}_k)/2$ corresponds to the intrinsic
parity projection of the correlated $NN$-pair 
in the $\alpha_k$ cluster.
Here, $\bvec{d}$ is chosen as 
$\bvec{d}=(d \sin\theta_\alpha, 0, d \cos\theta_\alpha)$, where
$d$ and $\theta_\alpha$ stands for the distance and angle of the 
relative position between the
two $\alpha$ clusters. 

\subsubsection{Full GCM calculation for two-$\alpha$ system}

Based on the generator coordinate method (GCM), all 
the AQCM-T wave function with
various $k$, $\theta_\alpha$, $\beta$, and $d$ values are superposed
as
\begin{eqnarray}\label{eq:gcm-8Be}
&&\Psi^\textrm{GCM}_{2\alpha,0^+}=\sum_{d}\Bigl[ c_0(d)
\Phi^{0s}_{2\alpha,0^+}(d) \nonumber\\
&& +
\sum_{k,\beta,\theta_\alpha} c(k,\beta,\theta_\alpha,d) 
\Phi^{\textrm{AQCM-T}}_{2\alpha,0^+}(k,\beta,\bvec{d})\Bigr],
\end{eqnarray}
where $\Phi^{0s}_{2\alpha,0^+}$ is the $(0s)^4$-$(0s)^4$ state with the distance $d$ 
given by the Brink-Bloch (BB)
wave function projected onto $J^\pi=0^+$ as, 
\begin{eqnarray}
&&\Phi^{0s}_{2\alpha,0^+}(d)=\widehat{P}^{0+}{\cal A}\left\{ 
\Phi_{\alpha_0}(\bvec{R}) 
\Phi_{\alpha_0}(\bvec{R}') 
\right\}.
\end{eqnarray}
In Eq.~(\ref{eq:gcm-8Be}),
the coefficients $c_0(d)$ and $c(k,\beta,\theta_\alpha,d)$ are determined 
by diagonalizing the norm and Hamiltonian matrices.
This is called full GCM calculation.

In addition to $k$ and 
$\beta$ of the correlated $NN$-pair in $\alpha_k$, the 
inter-cluster motion is described with the distance parameter $d$ and 
the angular
parameter $\theta_\alpha$.
It should be noted that the intrinsic wave function of  
$\alpha_k$ is axial symmetric and positive-parity state,
and that for the $\alpha_0$
is rotationally invariant, thus 
the angular range of
$0\le \theta_\alpha \le \pi/2$ is enough. 

\subsubsection{Fixed-$d$ calculation}\label{subsec:fixed-d}

In order to see properties of the two-$\alpha$ system
as a function of the inter-cluster distance $d$, 
we also show results with fixed value of $d$
called  ``fixed-$d$ calculation'',
\begin{eqnarray}\label{eq:fixed-d-gcm-8Be}
&&\Psi^\textrm{opt}_{2\alpha,0^+}(d)=f^{d}_0
\Phi^{0s}_{2\alpha,0^+}(d)\nonumber\\
&&+
\sum_{k,\beta,\theta_\alpha} f^{d}(k,\beta,\theta_\alpha) 
\Phi^{\textrm{AQCM-T}}_{2\alpha,0^+}(k,\beta,\bvec{d}).
\end{eqnarray}
For each $d$ value, 
the coefficients  $\{ f^{d}_0 \}$ and $\{ f^{d}(k,\beta,\theta_\alpha) \}$ are determined 
by diagonalizing the norm and Hamiltonian matrices. 
This means that the coefficients {$\{ f^{d}_0 \}$ and $\{ f^{d} \}$
are optimized so as to minimize the energy of the two-$\alpha$ system 
at each $d$,
\begin{equation}
E^\textrm{opt}(d)=\langle\Psi^\textrm{opt}_{2\alpha,0^+}(d)|\widehat{H}| \Psi^\textrm{opt}_{2\alpha,0^+}(d)\rangle,
\end{equation}
which corresponds to the adiabatic approximation.
To stress the optimization of the $\alpha$ cluster at each $d$, we also call the fixed-$d$ calculation ``optimized-$\alpha$''
calculation.

At large $\alpha$-$\alpha$ distances, the two-$\alpha$ system approaches the asymptotic state, in which 
each $\alpha$ cluster stays in the ground state as isolated $^4\textrm{He}$. 
This asymptotic state is approximately described by the fixed-$d$ calculation 
with large enough $d$. 
As two $\alpha$ clusters approach each other, each $\alpha$ cluster is excited because of the
Pauli blocking effect and potential energy effect from the other $\alpha$
cluster. 
The internal excitation 
of $\alpha$ clusters, which is usually called the core polarization, is taken into account by this
fixed-$d$ calculation in an adiabatic way. 

\subsubsection{Frozen-$\alpha$ calculation}\label{subsec:frozen}

We also show the 
energies of
``frozen $\alpha$-clusters'', where
the core polarization at short relative distances is omitted.
We use the $\alpha$ clusters obtained at the largest distance of the model, $d=d_\textrm{max}$,
for any value of $d$.
Namely, the
coefficients $f^{d_\textrm{max}}_0$ and $f^{d_\textrm{max}}(k,\beta,\theta_\alpha)$
are optimized at $d=d_\textrm{max}$ and
they are used for any $d$
as
 \begin{eqnarray}\label{eq:frozen-cluster-8Be}
&&\Psi^\textrm{frozen}_{2\alpha,0^+}(d)=f^{d_\textrm{max}}_0
\Phi^{0s}_{2\alpha,0^+}(d)\nonumber\\
&&+
\sum_{k,\beta,\theta_\alpha} f^{d_\textrm{max}}(k,\beta,\theta_\alpha) 
\Phi^{\textrm{AQCM-T}}_{2\alpha,0^+}(k,\beta,\bvec{d}).
\end{eqnarray}
The expectation values of the Hamiltonian at $d$ is then
\begin{equation}
E^\textrm{frozen}(d)=\langle\Psi^\textrm{frozen}_{2\alpha,0^+}(d)|\widehat{H}| 
\Psi^\textrm{frozen}_{2\alpha,0^+}(d)\rangle.
\end{equation}

\subsubsection{Internal and external energies}\label{subsec:internal-external}

We also estimate the contribution from the internal excitation of $\alpha$ cluster (internal energy) 
and the residual part (external energy) at each distance $d$. 
The energy difference between the optimized energy at $d$ and that at the largest distance 
of
$d=d_\textrm{max}$
is defined as
$E^\textrm{opt}_r(d)=E^\textrm{opt}(d)-E^\textrm{opt}(d_\textrm{max})$.
Here
$E^\textrm{opt}_r(d)$ contains
not only the effect of the internal excitation 
due to the core polarization ($\Delta E^\textrm{internal}$)
but also the external energy between clusters, which is
interpreted as a kind of $\alpha$-$\alpha$ potential 
where the internal excitation energy is excluded.
This means that $E_r^\textrm{opt}(d)$ can be divided into two parts as 
\begin{equation}
E_r^\textrm{opt}(d)=\Delta E^\textrm{internal}(d)+\Delta E^\textrm{external}(d).
\end{equation}
We estimate the internal excitation energies 
$\Delta E^\textrm{internal}(d)$ as follows.
We first perform the fixed-$d$ calculation at $d$.
Keeping the optimized coefficients at $d$ ($f^{d}_0$ and $f^{d}(k,\beta,\theta_\alpha)$), 
we change $d$ to the largest value, $d=d_\textrm{max}$, as
 \begin{eqnarray}\label{eq:cp-8Be}
&&\Psi^\textrm{core-exc}_{2\alpha,0^+}(d;d_\textrm{max})=f^{d}_0
\Phi^{0s}_{2\alpha,0^+}(d_\textrm{max})\nonumber\\
&&+
\sum_{k,\beta,\theta_\alpha} f^{d}(k,\beta,\theta_\alpha) 
\Phi^{\textrm{AQCM-T}}_{2\alpha,0^+}(k,\beta,\bvec{d}_\textrm{max}).
\end{eqnarray}
The corresponding energy is
\begin{equation}
E^\textrm{core-exc}(d)=\langle \Psi^\textrm{core-exc}_{2\alpha,0^+}(d;d_\textrm{max})|\widehat{H}| \Psi^\textrm{core-exc}_{2\alpha,0^+}(d;d_\textrm{max})\rangle.
\end{equation}
In this $\Psi^\textrm{core-exc}_{2\alpha,0^+}(d;d_\textrm{max})$, 
two $\alpha$ clusters are located with a large relative distance, but it still contains 
the core-polarization effect.
Therefore, the internal excitation energy due to the core-excitation can be evaluated 
by comparing it with 
the optimal energy at  $d_\textrm{max}$ as 
\begin{equation}
\Delta E^\textrm{internal}(d)\equiv E^\textrm{core-exc}(d)-E^\textrm{opt}(d_\textrm{max}).
\end{equation}
Then, we simply define the external energy as 
\begin{equation}
\Delta E^\textrm{external}(d)\equiv E^\textrm{opt}_r(d)-\Delta E^\textrm{internal}(d).
\end{equation}

\subsubsection{Parameter setting for $^8\textrm{Be}$}
When we calculate $^8\textrm{Be}$, 
we introduce truncated model spaces for each $\alpha$ cluster. 
As shown later, 
the results of the truncations reasonably reproduce 
the full calculation for $^4\textrm{He}$.

For the relative angle between the two $\alpha$ clusters, 
we adopt
four mesh points of $\theta_\alpha=(\pi/8) i$ ($i=1,\ldots,4$), 
which gives almost converged results. Although five points of 
$\theta_\alpha=(\pi/8) i$ ($i=0,\ldots,4$) were 
used in the previous paper~\cite{AQCM-T}, 
we omit $\theta_\alpha=0$,
which is less important, 
to reduce the computational cost. 
The energy difference of the calculations with and without $\theta_\alpha=0$
is $0.03$~MeV for the case of G3RS2-3R interaction at $d=6$~fm.

For the $\alpha$-$\alpha$ distance parameter $d$, 
eight values of $d=1,2,\ldots, 8$ fm are adopted.
This is bound state approximation and
$\alpha$ clusters are artificially confined 
in the range of $d\le d_\textrm{max} =8$ fm. 
Here $d_\textrm{max}=8$ fm is close to the position of the 
Coulomb barrier around $d\sim 7$ fm.

As a result, 
the number of states with 
the generator coordinates ($\theta_\alpha$, $k$, and $d$) and channel 
($\beta$) is 
200 (392) for the case of V2m-3R (G3RS2-3R) interaction.
The total number of the Slater determinants 
superposed 
is 584 (1032) for the V2m-3R (G3RS2-3R) case. 
This is called full GCM calculation for $^8$Be.

\subsection{$0s$, ${}^1S$, ${}^3S$, and ${}^3D$ probabilities}
In this study, we analyze the probabilities of the ${}^1S,{}^3S$, and ${}^3D$ channels,
\begin{eqnarray}
{\cal P}_{{}^1S,{}^3S,{}^3D}=|\langle \Psi|
\widehat{P}_{{}^1S,{}^3S,{}^3D} |\Psi \rangle|,
\end{eqnarray}
where $\widehat{P}_{{}^1S,{}^3S,{}^3D}$ are the projection operators onto the 
 ${}^1S,{}^3S$, and ${}^3D$ channels.
We also 
calculate the ``$0s$ probability''
\begin{eqnarray}
{\cal P}_{0s}=|\langle 0s|\Psi\rangle |^2,
\end{eqnarray}
where $|\Psi \rangle$ is the ground state wave function 
($|\Psi_{^4\textrm{He},\textrm{g.s.}} \rangle$ for $^4\textrm{He}$ and 
$|\Psi_{^8\textrm{Be},\textrm{g.s.}} \rangle$ for $^8\textrm{Be}$), 
and $| 0s \rangle$ is the state with the $(0s)^4$ configuration 
($ | \Phi^{0s}_{^4\textrm{He}} \rangle$ for $^4\textrm{He}$
and $ | \Phi^{0s}_{2\alpha,0^+} \rangle $ for $^8\textrm{Be}$).
The probabilities 
of the correlated ${}^1S$ and ${}^3S$ components orthogonal to the $0s$ state is calculated as
\begin{eqnarray}
{\cal P}^\perp_{{}^1S,{}^3S}&=&|\langle \Psi|\Lambda^\perp_{0s}
\widehat{P}_{{}^1S,{}^3S}\Lambda^\perp_{0s} | \Psi \rangle|^2,\\
\Lambda^\perp_{0s}&\equiv&1-|0s\rangle \langle 0s|.
\end{eqnarray}
Note that ${\cal P}^\perp_{{}^1S,{}^3S}$ somewhat depends on the adopted
width parameter $\nu$ of the $0s$ orbit.

\section{Hamiltonian and interactions}

\subsection{Hamiltonian}
The Hamiltonian used in the present calculation is
\begin{align}
\widehat{H}&=\sum_{i}^A
\widehat{T}_i-\widehat{T}_\mathrm{G}\nonumber\\
&
+\sum_{i<j}^A
\left[
\widehat{V}_\mathrm{c}(i,j)+\widehat{V}_\mathrm{ls}(i,j)+\widehat{V}_\mathrm{t}(i,j)+\widehat{V}_\mathrm{Coulomb}(i,j)\right],
\end{align}
where $\widehat{T}_i$ is the kinetic energy operator of $i$th nucleon. 
$\widehat{T}_\mathrm{G}$ is the total kinetic energy operator for the cm motion
and its expectation value,
$\langle \widehat{T}_\mathrm{G} \rangle=3\hbar\omega/4$ 
(= $3\hbar^2\nu/2m$), 
is constant in the present framework,
where $m$ is the mean value of proton and neutron masses. 

The two-body interaction consists of the central 
($\widehat{V}_\mathrm{c}$), spin-orbit ($\widehat{V}_\mathrm{ls}$),
tensor ($\widehat{V}_\mathrm{t}$), and 
Coulomb ($\widehat{V}_\mathrm{Coulomb}$) parts.
The Coulomb interaction for the protons is approximated by a seven-range Gaussian form.

Differently from our previous work~\cite{AQCM-T}, here we use
a realistic interaction, Gaussian soft-core potential with three ranges (G3RS interaction)~\cite{G3RS}, 
which reproduces the 
$NN$-scattering phase shifts. The G3RS interaction consists of the central part
with a soft core, spin-orbit part, and tensor part. 
The central part of G3RS has three-range Gaussian form as
\begin{align}
\widehat{V}_\mathrm{c}=&
\widehat{P}_{ij}(^3E)\sum_{n=1}^3V_{\mathrm{c},n}^{^3E}\exp\left(-\frac{r_{ij}^2}{\eta_{\mathrm{c},n}^2}\right)\nonumber\\
&+\widehat{P}_{ij}(^1E)\sum_{n=1}^3V_{\mathrm{c},n}^{^1E}\exp\left(-\frac{r_{ij}^2}{\eta_{\mathrm{c},n}^2}\right)\nonumber\\
&+\widehat{P}_{ij}(^3O)\sum_{n=1}^3V_{\mathrm{c},n}^{^3O}\exp\left(-\frac{r_{ij}^2}{\eta_{\mathrm{c},n}^2}\right)\nonumber\\
&+\widehat{P}_{ij}(^1O)\sum_{n=1}^3V_{\mathrm{c},n}^{^1O}\exp\left(-\frac{r_{ij}^2}{\eta_{\mathrm{c},n}^2}\right), 
\end{align}
where $\widehat{P}_{ij}(^{1,3}E)$ and  $\widehat{P}_{ij}(^{1,3}O)$ are the projection operators to the 
$^{1,3}E$ (singlet-even, triplet-even) and $^{1,3}O$  (singlet-even, triplet-odd) states, respectively. 
We simply call the even (odd) part of the central interaction the 
central-even (central-odd) interaction.

The spin-orbit part of G3RS has a two-range Gaussian form as
\begin{align}\label{eq:ls}
\widehat{V}_\mathrm{ls}=&\widehat{\bvec{L}}_{ij}\cdot\widehat{\bvec{S}}_{ij}\times
\nonumber\\
&\widehat{P}_{ij}(^3O)\sum_{n=1}^{n_\textrm{max}=3}
V_{\mathrm{ls},n}^{^3O}
\exp\left(-\frac{r_{ij}^2}{\eta_{\textrm{ls},n}^2}\right).
\end{align}

For the tensor part, 
we use the version,
where the G3RS tensor part
is fitted with the $r^2$-weighted Gaussian form
(3-range fit tensor)
given in the previous paper~\cite{AQCM-T},

\begin{align}
\widehat{V}_\mathrm{t}=&r_{ij}^2\widehat{S}_{ij}\times\nonumber\\
&\left[ 
\widehat{P}_{ij}(^3E)\sum_{n=1}^{n_\textrm{max}=3}
V_{\mathrm{t},n}^{^3E}
\exp\left(-\frac{r_{ij}^2}{\eta_{\textrm{t},n}^2}\right)\right. \nonumber\\
&+\left. 
\widehat{P}_{ij}(^3O)\sum_{n=1}^{n_\textrm{max}=3}
V_{\mathrm{t},n}^{^3O}
\exp\left(-\frac{r_{ij}^2}{\eta_{\textrm{t},n}^2}\right)\right],\\
r_{ij}^2\widehat{S}_{ij}=&3(\widehat{\bvec{\sigma}}_i\cdot\widehat{\bvec{r}}_{ij})(\widehat{\bvec{\sigma}}_j\cdot\widehat{\bvec{r}}_{ij})
-(\widehat{\bvec{\sigma}}_i\cdot\widehat{\bvec{\sigma}}_j)r_{ij}^2,\nonumber\\
\label{eq:tensor}
\end{align}
where the standard tensor operator $\widehat{S}_{ij}$ is dimensionless, 
and the range parameters  $\eta_{{\textrm{t},n}}$ 
is 
related to the 
width parameters $\beta_n$ in the previous paper~\cite{AQCM-T}
as
$\beta_n=1/\eta^2_{{\textrm{t},n}}$.
The accuracy of the fitting is
rather well within a few percent errors in energy 
for the ground state of $^4\textrm{He}$. 

\subsection{Interaction parameters}

We use ``case 1'' and ``case 2" parameters of G3RS, which are labeled as G3RS1 and 
G3RS2. 
The central part of G3RS1 has 
the repulsive core 
of 2000~MeV height for
the even-parity state, which is designed 
to reproduce
 the $NN$-scattering phase shift
up to 600~MeV, 
whereas G3RS2 has the 500-630 MeV height and reproduces the $NN$-scattering up
to 150~MeV. 
As mentioned before, the tensor part is approximated by the 3-range fit and therefore we
note the case 1 and case 2 interactions
as G3RS1-3R and G3RS2-3R, respectively.
For comparison, we also use an effective interaction, 
``V2m-3R'' introduced in the previous paper~\cite{AQCM-T},
which explicitly includes the tensor part 
but no repulsive core for the central part.
Here the $^3E$ part of the central interaction (Volkov No.2~\cite{Volkov}) is
reduced to 60\% of the original strength so as to reproduce the correct
binding energy of $^4\textrm{He}$ after adding the tensor term 
within the AQCM-T model space. 
The spin-orbit part of V2m-3R is the same as G3RS1, 
and the tensor part is 3-range fit as in G3RS1-3R and G3RS2-3R.
As shown in the next section,  
the binding energy and radius of $^4\textrm{He}$
are reasonably reproduced 
with G3RS2-3R and V2m-3R, but not 
satisfactory with G3RS1-3R within the present AQCM-T framework
because of the high repulsive core. 
Therefore, we mainly use
G3RS2-3R as the default interaction and compare the result with V2m-3R. 

We also show 
results of conventional $\alpha$-cluster model, 
where $\alpha$ clusters are expressed as pure $(0s)^4$ configurations 
without the $NN$ correlation.
Here effective interactions with no repulsive core for the central part
and no tenser terms are used;
the Volkov interaction with
the same parameter as in the  previous paper~\cite{AQCM-T}, 
labeled as ``V2''.
Note that here only the central interaction gives non-zero 
contribution, and the spin-orbit and tensor contributions vanish 
even if the interaction contains such spin-dependent terms
because 
the intrinsic spins are saturated in the $(0s)^4$ configuration. 
Even though V2 well describes basic properties of very light-mass nuclei, it is
not sufficient for the saturation property of the nuclear matter;
it causes the overbinding problem in
heavier nuclei 
($A\ge 16$).
To avoid this problem, the odd-parity part has to be tuned 
for heavier nuclei by modifying the Majorana parameter as often done in 
conventional cluster model calculations.
The Majorana parameter of V2 used here is adjusted to explain the 
$\alpha$-$\alpha$ scattering phase shift.

We also show results obtained with the Brink-Boeker interaction, 
which does not have the tensor term
but gives the nuclear saturation.
We use ``case four" of the Brink-Boeker interaction labeled as BB4~\cite{BB}.
This interaction has
a similar form to V2  (two range Gaussians) but it has a quite shorter range for
the odd part compared with V2 as 
$\eta_{c,2}=0.4$~fm, which is almost a contact interaction. 
The BB4 interaction was designed
to reproduce the  energy and density of the nuclear matter 
at the saturation point as well as the binding energy of $^4\textrm{He}$, but instead, 
it cannot reproduce the $NN$ and $\alpha$-$\alpha$ scattering 
phase shifts. It also tends to give less binding energies for 
closed shell nuclei such as $^{16}$O and $^{40}$Ca.

In Table \ref{table:int},  we summarize the parameters of G3RS1-3R, G3RS2-3R, V2m-3R, V2, and BB4.
It should be commented again 
that the same 3-range fit tensor is used in these three cases
(G3RS1-3R, G3RS2-3R, and V2m-3R).
{The $^1E$ part 
of V2m-3R is the same as that of V2, which is
adjusted to reproduce the $NN$ scattering length
of the $^1S$ channel. On the other hand, 
the $^3E$ part of V2m-3R is  
reduced to
60\% of that of V2 so as to reasonably describe the 
binding energy and radius of $^4\textrm{He}$ after including the tensor term. 
More details on the V2m-3R and V2 interactions are explained in our previous paper~\cite{AQCM-T}.
In Fig.~\ref{fig:pot}, we show 
the $r$ dependence
for the central and tensor parts of 
G3RS1-3R, G3RS2-3R, and V2m-3R
for the $^3E$  channel. 
For the central part, 
the realistic interactions, G3RS1 ($V_\textrm{c}$:G3RS1)
and G3RS2 
($V_\textrm{c}$:G3RS2),
have the short-range repulsion and 
middle-range attraction.
As clearly seen,
G3RS1
has 
higher core and deeper pocket than G3RS2, 
whereas an effective interaction,
the central part of V2m-3R ($V_\textrm{c}$:V2m), has 
no repulsive core and the attractive part has
longer range.
For the tensor part,
here the original form of the G3RS2 interaction
($V_\textrm{t}$:G3RS2) and its 3-range fit ($V_\textrm{t}$:3R) are presented.

\begin{table*}[!thbp]
\caption{The parameter sets of G3RS1-3R, 
G3RS2-3R, V2m-3R, V2, and BB4
interactions.
The range parameters of the tensor interaction, $\eta_{{\textrm{t},n}}$,
is 
related to the 
width parameters $\beta_n$ in the previous paper~\cite{AQCM-T}
as
$\beta_n=1/\eta^2_{{\textrm{t},n}}$, 
and
$\beta_{1,2,3} = 0.53, 1.92, 8.95$~fm$^{-2}$.
\label{table:int}
}
\centering
\begin{tabular}{lrrrrr|lrrr}
\hline \hline
&\multicolumn{3}{c}{central} & \multicolumn{2}{c|}{ls}& &\multicolumn{3}{c}{tensor}\\ 
$n$	&$	1	$&$	2	$&$	3	$&$	1	$&$	2	$&$	 $& $1	$&$	2	$&$	3	$\\
\hline
\multicolumn{6}{l|}{G3RS1-3R} \\
$\eta_{\textrm{c/ls},n}$ (fm)	&$	2.5	$&$	0.942	$&$	0.447	$&$	0.6	$&$	0.447	$&	$\eta_{\textrm{t},n}$ (fm)	&$	1.3736 	$&$	0.7217 	$&$	0.3343 	$\\
	
$V_{\textrm{c/ls},n}^{^3E}$ (MeV)	&$	-5	$&$	-210	$&$	2000	$&$		$&$		$&	$V_{\textrm{t},n}^{^3E}$ (MeV fm$^{-2}$)	&$	-17.02	$&$	-209.89	$&$	-289.59	$\\
$V_{\textrm{c/ls},n}^{^1E}$ (MeV)	&$	-5	$&$	-270	$&$	2000	$&$		$&$		$&		&$		$&$		$&$		$\\
$V_{\textrm{c/ls},n}^{^3O}$ (MeV)	&$	1.6667	$&$	-50	$&$	2500	$&$	-1050	$&$	600	$&	$V_{\textrm{t},n}^{^3O}$ (MeV fm$^{-2}$)	&$	5.27	$&$	62.91	$&$	89.87	$\\
$V_{\textrm{c/ls},n}^{^1O}$ (MeV)	&$	10	$&$	50	$&$	2000	$&$		$&$		$&		&$		$&$		$&$		$\\
\hline
\multicolumn{6}{l|}{G3RS2-3R} \\
$\eta_{\textrm{c/ls},n}$ (fm)	&$	2.5	$&$	0.942	$&$	0.6	$&$	0.6	$&$	0.4	$&	$\eta_{\textrm{t},n}$ (fm)	&$	1.3736 	$&$	0.7217 	$&$	0.3343 	$\\
$V_{\textrm{c/ls},n}^{^3E}$ (MeV)	&$	-5	$&$	-210	$&$	500	$&$		$&$		$&	$V_{\textrm{t},n}^{^3E}$ (MeV fm$^{-2}$)	&$	-17.02	$&$	-209.89	$&$	-289.59	$\\
$V_{\textrm{c/ls},n}^{^1E}$ (MeV)	&$	-5	$&$	-270	$&$	630	$&$		$&$		$&		&$		$&$		$&$		$\\
$V_{\textrm{c/ls},n}^{^3O}$ (MeV)	&$	1.6667	$&$	-50	$&$	400	$&$	-800	$&$	800	$&	$V_{\textrm{t},n}^{^3O}$ (MeV fm$^{-2}$)	&$	5.27	$&$	62.91	$&$	89.87	$\\
$V_{\textrm{c/ls},n}^{^1O}$ (MeV)	&$	10	$&$	50	$&$	200	$&$		$&$		$&		&$		$&$		$&$		$\\
\hline
\multicolumn{6}{l|}{V2m-3R} \\
$\eta_{\textrm{c/ls},n}$ (fm)	&$	1.8	$&$	1.01	$&$		$&$	0.6	$&$	0.447	$&	$\eta_{\textrm{t},n}$ (fm)	&$	1.3736 	$&$	0.7217 	$&$	0.3343 	$\\
$V_{\textrm{c/ls},n}^{^3E}$ (MeV)	&$	-47.307 	$&$	47.689 	$&$		$&$		$&$		$&	$V_{\textrm{t},n}^{^3E}$ (MeV fm$^{-2}$)	&$	-17.02	$&$	-209.89	$&$	-289.59	$\\
$V_{\textrm{c/ls},n}^{^1E}$ (MeV)	&$	-42.455	$&$	42.798	$&$		$&$		$&$		$&		&$		$&$		$&$		$\\
$V_{\textrm{c/ls},n}^{^3O}$ (MeV)	&$	12.13	$&$	-12.228	$&$		$&$	-1050	$&$	600	$&	$V_{\textrm{t},n}^{^3O}$ (MeV fm$^{-2}$)	&$	5.27	$&$	62.91	$&$	89.87	$\\
$V_{\textrm{c/ls},n}^{^1O}$ (MeV)	&$	12.13	$&$	-12.228	$&$		$&$		$&$		$&		&$		$&$		$&$		$\\
\hline
\multicolumn{6}{l|}{V2} \\
$\eta_{\textrm{c},n}$ (fm)	&$	1.8	$&$	1.01	$&$		$&$$&$	$&$		$&$		$&$		$&$		$\\
$V_{\textrm{c},n}^{^3E}$ (MeV)	&$	-78.845	$&$	79.482	$&$		$&$		$&$		$&$		$&$		$&$		$&$		$\\
$V_{\textrm{c},n}^{^1E}$ (MeV)	&$	-42.455	$&$	42.798	$&$		$&$		$&$		$&$		$&$		$&$		$&$		$\\
$V_{\textrm{c},n}^{^3O}$ (MeV)	&$	12.13	$&$	-12.228	$&$		$&$		$&$	$&$		$&$		$&$		$&$		$\\
$V_{\textrm{c},n}^{^1O}$ (MeV)	&$	12.13	$&$	-12.228	$&$		$&$		$&$		$&$		$&$		$&$		$&$		$\\
\hline
\multicolumn{6}{l|}{BB4} \\
$\eta_{\textrm{c},n}$ (fm)	&$	0.8	$&$	0.4	$&$		$&$		$&$		$&		&$		$&$		$&$		$\\
$V_{\textrm{c},n}^{^3E}$ (MeV)	&$	-1307.9	$&$	7228	$&$		$&$		$&$		$&		&$		$&$		$&$		$\\
$V_{\textrm{c},n}^{^1E}$ (MeV)	&$	-1307.9	$&$	7228	$&$		$&$		$&$		$&		&$		$&$		$&$		$\\
$V_{\textrm{c},n}^{^3O}$ (MeV)	&$	-60.4 	$&$	7097.9 	$&$		$&$		$&$		$&		&$		$&$		$&$		$\\
$V_{\textrm{c},n}^{^1O}$ (MeV)	&$	-60.4 	$&$	7097.9 	$&$		$&$		$&$		$&		&$		$&$		$&$		$\\
\hline\hline
\end{tabular}
\end{table*}

\begin{figure}[!thb]
\begin{center}
\includegraphics[width=8.5cm]{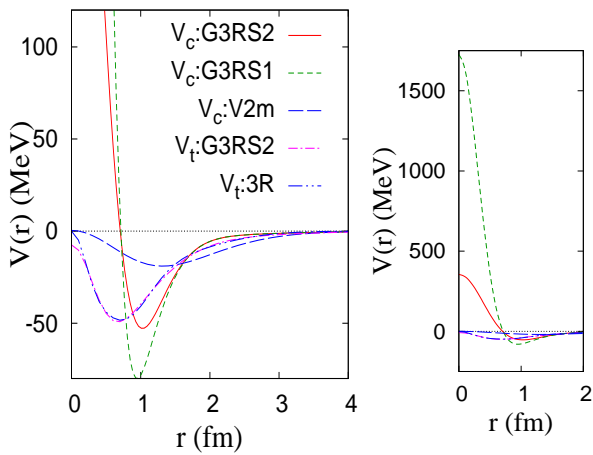} 
\end{center}
  \caption{\label{fig:pot}
 (Color online) 
Left: radial ($r$) dependence of the $^3E$ parts of the central ($V_\textrm{c}$) and tensor  ($V_\textrm{t}$) interactions 
of G3RS1-3R, G3RS2-3R, and V2m-3R. 
The 3-range fit tensor is used in these three cases. 
Right: same as the left panel but on a different scale.}  
\end{figure}

\section{Application to $^4\textrm{He}$} \label{sec:results-he4}
We apply the AQCM-T method to $^4$He using G3RS1-3R, G3RS2-3R, and V2m-3R.
For comparison, we also show the results obtained with V2,
where the model space is conventional cluster model
with the $(0s)^4$ configuration, which we call the V2:$0s$ calculation.

\subsection{Results of  $^4$He}

\begin{table}[!tbh]
\caption{Energies, radii $(R_m)$, and probabilities $({\cal P})$ of $^4\textrm{He}$
obtained by AQCM-T full configurations. 
For energies, the expectation values of the total energy $(E)$, kinetic energy $(T)$, central interaction $(V_\textrm{c})$, and tensor interaction $(V_\textrm{t})$ are shown. 
The calculated results
for V2m-3R, G3RS2-3R, and G3RS1-3R are listed. 
The result of V2 with the $(0s)^4$ configuration is also shown (V2:$0s$).
The experimental values of the total energy and radius are 
$E=-28.296$~MeV and $R_m=1.455$~fm \cite{angeli13}. 
 \label{tab:energy}
}
\begin{center}
\begin{tabular}{cccccccccc}
\hline
\hline
	&	V2m-3R	&	G3RS2-3R	&	G3RS1-3R	&	V2:$0s$	\\	
$E$ (MeV)	&$	-30.3 	$&$	-26.5 	$&$	-16.2 	$&$	-27.9 	$\\	
$T$ (MeV)	&$	64.6 	$&$	72.3 	$&$	70.9 	$&$	46.7 	$\\	
$V_\textrm{c}$ (MeV)	&$	-56.7 	$&$	-58.4 	$&$	-54.6 	$&$	-75.3 	$\\	
$V_\textrm{t}$ (MeV)	&$	-39.9 	$&$	-41.7 	$&$	-33.9 	$&$		$\\	
$R_m$ (fm)	&$	1.46 	$&$	1.43 	$&$	1.53 	$&$	1.50 	$\\	
${\cal P}_{0s}$	&$	0.901 	$&$	0.881 	$&$	0.897 	$&$		$\\	
${\cal P}_{{}^3D}$	&$	0.077 	$&$	0.082 	$&$	0.063 	$&$		$\\	
${\cal P}^\perp_{{}^3S}$	&$	0.018 	$&$	0.028 	$&$	0.022 	$&$		$\\	
${\cal P}^\perp_{{}^1S}$	&$	0.004 	$&$	0.015 	$&$	0.020 	$&$		$\\	
\hline
\hline
\end{tabular}
\end{center}
\end{table}

The total energy ($E$), contributions of the
kinetic term ($T$),  central ($V_\textrm{c}$) and tensor ($V_\textrm{t}$) interactions,  
root-mean-square (rms) matter radius ($R_m$), 
and the probabilities of $0s$, ${}^1S$, ${}^3S$, and ${}^3D$ channel 
of $^4\textrm{He}$
are listed in Table~\ref{tab:energy}.
The results calculated with  G3RS2-3R and V2m-3R reasonably 
reproduces the experimental binding energy and radius.
For the energy contribution of each term ($T$, $V_\textrm{c}$, $V_\textrm{t}$),  
the two interactions, G3RS2-3R and V2m-3R, give similar results even though only the
former has a short-range repulsion
in the central part. 
They also give similar values for the $D$-state probability (${\cal P}_{^3D}$) meaning 
that the tensor contributions  
are qualitatively the same in both of them. 
On the other hand, 
for the $S$-wave,
G3RS2-3R shows a 
significant mixing of the correlated ${^1S}$-state
(${\cal P}^\perp_{^1S}$) and enhancement of the $^3S$ 
 correlations (${\cal P}^\perp_{^3S}$);
the correlated $^1S$ and  $^3S$ pairs 
 beyond the $(0s)^4$ configuration are important
owing to the short-range core of the central part.
This result confirms that V2m-3R is an effective interaction,
which 
explicitly includes the tensor interaction but the repulsive core  
in the central part is renormalized.

The results of G3RS1-3R are qualitatively similar 
to G3RS2-3R,
but quantitatively different; 
the binding energy ($-E$) 
is underestimated compared with the experiment,
whereas the size {$(R_m)$
is overestimated
because of the higher repulsive core
of G3RS1-3R (the height is 2000~MeV and  $500\sim 630$~MeV in G3RS1 and G3RS2
in the central part, respectively).

These results indicate that the present AQCM-T method is applicable 
for a kind of {\it ab initio} calculation of $^4\textrm{He}$.
Although the repulsive core
of G3RS2 is lower than G3RS1, G3RS2 is 
still 
a ``realistic force'',  which reproduces 
the phase shift of low-energy $NN$ scattering up to $\sim$150 MeV. 
However, the present model space is not sufficient for G3RS1-3R, a realistic force with 
a significant height of the core capable of reproducing the $NN$ scattering
of higher energies.

Hereafter, we mainly focus on the results of G3RS2-3R 
and discuss features of the tensor and short-range correlations in $^4\textrm{He}$
and their roles in the two-$\alpha$ system, while comparing them with V2m-3R and V2:$0s$.

\subsection{Contribution of each basis states in $^4\textrm{He}$}

\begin{figure}[!htb]
\begin{center}
\includegraphics[width=6cm]{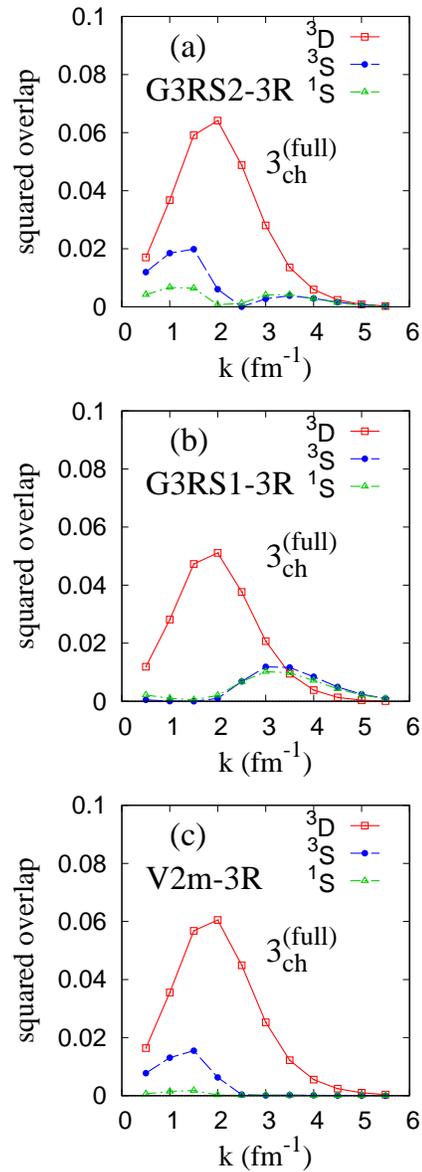} 
\end{center}
  \caption{	\label{fig:k}
 (Color online) 
Squared overlaps ${\cal O}_\beta(k)$ for the $\beta={}^1S$, ${}^3S$, and ${}^3D$ channels 
of $^4\textrm{He}(\textrm{g.s.})$ obtained with the full configurations ($3^\textrm{(full)}_\textrm{ch}$).
The results for (a) G3RS2-3R, (b) G3RS1-3R, and (c) V2m-3R are shown.}
\end{figure}

To clarify which channel ($\beta$) and momentum region ($k$)}
contribute to the tensor and short-range correlations 
in $^4$He, we first 
analyze the squared overlap between 
the 
ground state wave function $\Psi_{^4\textrm{He},\textrm{g.s.}}$
and each basis state of AQCM-T.
Each basis state is 
specified by the momentum parameter $k$ and the channel $\beta$;
the squared overlap (${\cal O}_\beta(k)$)
for $k$ in the $\beta={}^3D$ channel
is calculated as 
\begin{eqnarray}
{\cal O}_{{}^3D}(k)=
|\langle \Phi^{\textrm{AQCM-T}}_{^4\textrm{He},0^+}(k,{}^3D)|
\Psi_{^4\textrm{He},\textrm{g.s.}}
\rangle|^2.
\end{eqnarray}
The squared overlaps for the $\beta={}^{1,3}S$ channels are 
defined for their  components orthogonal to $|0s\rangle$ as 
\begin{eqnarray}
{\cal O}_{{}^{1,3}S}(k)=
|\langle \Phi^{\textrm{AQCM-T}}_{^4\textrm{He},0^+}(k,{}^{1,3}S)\Lambda^\perp_{0s}|
\Psi_{^4\textrm{He},\textrm{g.s.}}
\rangle|^2, 
\end{eqnarray}
which measure the correlated $^{1,3}S$ components beyond 
the simple $0s$ (uncorrelated) state.

As shown in Fig.~\ref{fig:k},
in the two cases of G3RS2-3R (Fig.~\ref{fig:k}~(a))
and V2m-3R  (Fig.~\ref{fig:k}~(c)), 
the $^3D$ pair has
the largest overlap
around $k\sim 2$~fm$^{-1}$, indicating that 
this region of $k$ dominantly contributes to the 
tensor correlation. 
This result is consistent with that of our previous paper~\cite{AQCM-T} and also
qualitatively in agreement
with the discussion by Myo {\it et al.} \cite{hmAMD}, which pointed out
the importance of the high-momentum tensor correlation.
For the $^3S$ and $^1S$ channels,
both of
G3RS2-3R (Fig.~\ref{fig:k}~(a))
and  V2m-3R (Fig.~\ref{fig:k}~(c))
contain the components 
in the lower momentum region of
$k < 2$~fm$^{-1}$, but
one can see a  
striking difference in the high momentum regions;  
G3RS2-3R contains 
high momentum 
components 
of $k=3\sim 4$~fm$^{-1}$ because of the short-range correlation caused by the repulsive core
of the central interaction, but 
V2m-3R does not.

Let us turn to G3RS1-3R  (Fig.~\ref{fig:k}~(b)). The $^3D$ pair has the largest overlap 
around $k\sim 2$~fm$^{-1}$, consistently with G3RS2-3R and V2m-3R. 
It also contains high momentum components of the $^3S$ and $^1S$ channels
in the 
$k=3\sim 4$~fm$^{-1}$ region caused by the short-range core,
which are even enhanced compared with G3RS2-3R.

\subsection{Truncation of the model space for $^4\textrm{He}$}

\begin{figure}[!htb]
\begin{center}
\includegraphics[width=7cm]{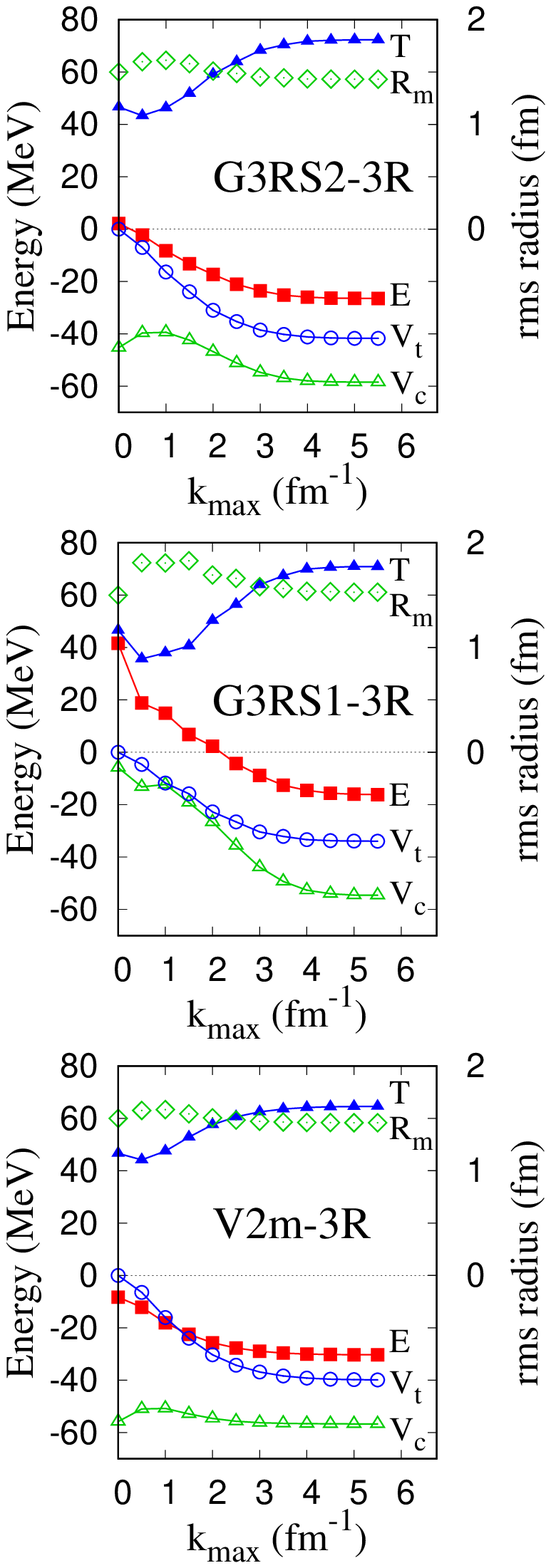} 
\end{center}
  \caption{	\label{fig:kmax} (Color online) 
Energies (left vertical axis) and radii (right vertical axis) of $^4\textrm{He}$
for the
3-channel calculation ($3^{k\le k_\textrm{max}}_\textrm{ch}$)
with the truncation for the momentum $k$ ($k\le k_\textrm{max}$).
For energies, the expectation values of the total energy $(E)$, kinetic energy $(T)$, central interaction $(V_\textrm{c})$, and tensor interaction $(V_\textrm{t})$ are shown. The radii $(R_m)$ are shown by disconnected diamonds. 
The upper-, middle-, and lower panels show the results of
G3RS2-3R,  G3RS1-3R, and V2m-3R, respectively.
}
\end{figure}

Next, we perform the AQCM-T calculation 
with truncated model space, where the configurations are reduced from 
the full calculation.
In Fig.~\ref{fig:kmax}, we show the results of three-channel calculation with truncated 
configurations for $k$ as  $k\le k_\textrm{max}$ (labeled by $3^{k\le k_\textrm{max}}_\textrm{ch}$). 
The total energy ($E$), the contribution of each term of the Hamiltonian
($T$, $V_\textrm{c}$, $V_\textrm{t}$), 
and rms radius  ($R_m$)
are shown for
G3RS2-3R (upper panel),  G3RS1-3R (middle panel), and V2m-3R (lower panel).
In V2m-3R (lower panel), the contribution of the central term ($V_\textrm{c}$)
is almost unchanged 
in spite of the reduction of the $k$ values 
and it stays at the value close to the one for the $(0s)^4$ configuration.
The kinetic energy  ($T$) gradually 
increases with increasing $k_\textrm{max}$ in the range of 
$k_\textrm{max}\lesssim 3$ fm$^{-1}$, whereas
the tensor interaction ($V_\textrm{t}$) compensates or even overcomes it. 
As a result, the total energy ($E$) gradually decreases and
almost converges around $k_\textrm{max}\gtrsim 3$~fm$^{-1}$.
This compensation of increasing kinetic energy and attractive tensor energy
is the typical feature when the $^3D$ component mixes, which can be also 
seen in G3RS1-3R (Fig.~\ref{fig:kmax}~(b)) 
and G3RS2-3R (Fig.~\ref{fig:kmax}~(a)). 
In G3RS2-3R (Fig.~\ref{fig:kmax}~(a)), 
the $k_\textrm{max}$ dependence of the tensor 
($V_\textrm{t}$) 
and total 
{($E$)
energies 
are qualitatively similar to those of V2m-3R (Fig.~\ref{fig:kmax}~(c)); however, a clear difference is found 
in the central part
($V_\textrm{c}$), which gradually decreases 
in the 1~fm$^{-1}\lesssim k_\textrm{max} \lesssim 4$~fm$^{-1}$ region.
Another difference is a rather rapid increase of the kinetic energy ($T$)
seen in G3RS2-3R 
compared with V2m-3R.
These behaviors of G3RS2-3R are qualitatively similar to G3RS1-3R,  but
quantitatively, 
the $k_\textrm{max}$ dependence of the total, kinetic and central energies is stronger in G3RS1-3R. 
From these analyses, we can conclude that   
$k\lesssim 3$~fm$^{-1}$ is essential for the tensor correlation, while 
$k\lesssim 4$~fm$^{-1}$ contributes to the short-range correlation.

\begin{table*}[!htbp]
\caption{Energies, radii $(R_m)$, and probabilities $({\cal P})$ of $^4\textrm{He}$
obtained with AQCM-T
are compared with the $(0s)^4$ configuration.
For energies, 
 the expectation values of the total energy $(E)$, kinetic energy $(T)$, 
central interaction $(V_\textrm{c})$, and tensor interaction $(V_\textrm{t})$ are shown. 
The results of $3^\textrm{(full)}_\textrm{ch}$, 
$2^\textrm{(full)}_\textrm{ch}$, and $1^\textrm{(full)}_\textrm{ch}$ 
calculations with full $k$ values ($k=\{0.5, 1.0, \ldots, 5.5\}$ fm$^{-1}$)
for G3RS2-3R and V2m-3R are listed.
The results of $3^\textrm{(4)}_\textrm{ch}$ 
(three channels with four $k$ values, $k=\{1.5, 2.5, 3.5, 4.5\}$ fm$^{-1}$),
$3^\textrm{(3)}_\textrm{ch}$ (three channels with 
three $k$ values, $k=\{1.5, 2.5, 4.0\}$ fm$^{-1}$), 
$2^\textrm{(3)}_\textrm{ch}$ (two channels with three $k$ values, $k=\{1.0, 2.0, 3.0\}$ fm$^{-1}$) are also shown.
For the results of G3RS2-3R, 
the expectation values of the long-range ($V_\textrm{c,1}$:$\eta_{c,1}=$2.5 fm), middle-range ($V_\textrm{c,2}$:$\eta_{c,2}=$0.942 fm), and 
short-range ($V_\textrm{c,3}$:$\eta_{c,3}=$0.6 fm) terms of the central interactions are also shown. 
\label{tab:truncation}
}
\begin{center}
\begin{tabular}{c|cccc|ccccc}
\hline
\hline
\multicolumn{6}{c}{G3RS2-3R}	\\									
	&	3$^\textrm{(full)}_\textrm{ch}$	&	2$^\textrm{(full)}_\textrm{ch}$	&	1$^\textrm{(full)}_\textrm{ch}$	&	$(0s)^4$	&	$3^\textrm{(4)}_\textrm{ch}$	&	$3^\textrm{(3)}_\textrm{ch}$	\\
$\beta$	&	$\{^1S,^3S,^3D\}$	&	$\{^3S,^3D\}$	&	$\{^3D\}$	&		&	$\{^1S,^3S,^3D\}$	&	$\{^1S,^3S,^3D\}$	\\
$k$	&	full	&	full	&	full	&		&	$\{1.5,2.5,3.5,4.5\}$	&	$\{1.5,2.5,4.0\}$	\\
$E$ (MeV)	&$	-26.5 	$&$	-22.7 	$&$	-18.5 	$&$	2.2 	$&$	-26.0 	$&$	-25.1 	$\\
$T$ (MeV)	&$	72.3 	$&$	68.9 	$&$	59.8 	$&$	46.7 	$&$	73.6 	$&$	72.9 	$\\
$V_\textrm{c}$ (MeV)	&$	-58.4 	$&$	-51.2 	$&$	-43.6 	$&$	-45.3 	$&$	-59.1 	$&$	-58.4 	$\\
$V_\textrm{t}$ (MeV)	&$	-41.7 	$&$	-41.9 	$&$	-36.2 	$&$		$&$	-41.8 	$&$	-41.0 	$\\
$V_\textrm{c,1}$ (MeV)	&$	-14.6 	$&$	-14.3 	$&$	-13.6 	$&$	-14.3 	$&$	-14.8 	$&$	-14.7 	$\\
$V_\textrm{c,2}$ (MeV)	&$	-124.1 	$&$	-123.3 	$&$	-108.7 	$&$	-111.4 	$&$	-125.5 	$&$	-124.5 	$\\
$V_\textrm{c,3}$ (MeV)	&$	80.3 	$&$	86.5 	$&$	78.7 	$&$	80.4 	$&$	81.2 	$&$	80.8 	$\\
$R_m$ (fm)	&$	1.43 	$&$	1.44 	$&$	1.49 	$&$	1.50 	$&$	1.41 	$&$	1.41 	$\\
${\cal P}_{0s}$	&$	0.881 	$&$	0.891 	$&$	0.919 	$&$	1.000 	$&$	0.885 	$&$	0.886 	$\\
${\cal P}_{{}^3D}$	&$	0.082 	$&$	0.085 	$&$	0.081 	$&$		$&$	0.081 	$&$	0.080 	$\\
${\cal P}^\perp_{{}^3S}$	&$	0.028 	$&$	0.024 	$&$		$&$		$&$	0.027 	$&$	0.026 	$\\
${\cal P}^\perp_{{}^1S}$	&$	0.015 	$&$		$&$		$&$		$&$	0.015 	$&$	0.015 	$\\
\hline
\multicolumn{6}{c}{V2m-3R}	\\
	&	3$^\textrm{(full)}_\textrm{ch}$	&	2$^\textrm{(full)}_\textrm{ch}$	&	1$^\textrm{(full)}_\textrm{ch} $	&	$(0s)^4$	&	$2^\textrm{(3)}_\textrm{ch}$	\\	
$\beta$	&	$\{^1S,^3S,^3D\}$	&	$\{^3S,^3D\}$	&	$\{^3D\}$	&	&	$\{^3S,^3D\}$		\\	
$k$	&	full	&	full	&	full	&  &			$\{1.0,2.0,3.0\}$	\\	
$E$ (MeV)	&$	-30.3 	$&$	-30.0 	$&$	-28.2 	$&$	-8.3 	$&$	-28.2 	$\\	
$T$ (MeV)	&$	64.6 	$&$	65.7 	$&$	58.9 	$&$	46.7 	$&$	64.8 	$\\	
$V_\textrm{c}$ (MeV)	&$	-56.7 	$&$	-57.1 	$&$	-53.7 	$&$	-55.8 	$&$	-57.4 	$\\	
$V_\textrm{t}$ (MeV)	&$	-39.9 	$&$	-40.2 	$&$	-35.0 	$&$		$&$	-37.3 	$\\	
$R_m$ (fm)	&$	1.46 	$&$	1.44 	$&$	1.49 	$&$	1.50 	$&$	1.43 	$\\	
${\cal P}_{0s}$	&$	0.901 	$&$	0.903 	$&$	0.927 	$&$	1.00 	$&$	0.905 	$\\	
${\cal P}_{{}^3D}$	&$	0.077 	$&$	0.078 	$&$	0.073 	$&$		$&$	0.080 	$\\	
${\cal P}^\perp_{{}^3S}$	&$	0.018 	$&$	0.019 	$&$		$&$		$&$	0.016 	$\\	
${\cal P}^\perp_{{}^1S}$	&$	0.004 	$&$		$&$		$&$		$&$		$\\	
\hline
\hline
\end{tabular}
\end{center}
\end{table*}

We perform further analysis based on the truncation of  channels $\beta=\{^1S,^3S,^3D\}$;
the two-channel calculation with $\beta=\{^3S,^3D\}$ and the single-channel calculation 
only with $\beta=\{^3D\}$ 
are performed, while $k$ values are not truncated.
They are labeled as $3^\textrm{(full)}_\textrm{ch}$, $2^\textrm{(full)}_\textrm{ch}$, and 
$1^\textrm{(full)}_\textrm{ch}$, respectively.       
In Table \ref{tab:truncation},
the contribution of each term of the Hamiltonian, radius, and 
probabilities in the three cases ($3^\textrm{(full)}_\textrm{ch}$, $2^\textrm{(full)}_\textrm{ch}$,
and  $1^\textrm{(full)}_\textrm{ch}$) for G3RS2-3R and V2m-3R are shown together with those for the $(0s)^4$ 
configuration. 
In the case of G3RS2-3R, 
the only inclusion of the $^3D$ channel ($1^\textrm{(full)}_\textrm{ch}$)
describes major part of the tensor correlation. 
However, 
as seen in the significant difference between $1^\textrm{(full)}_\textrm{ch}$ 
and $2^\textrm{(full)}_\textrm{ch}$, 
the mixing of the correlated $^3S$ component also somewhat contributes
to the energy gains of the central ($V_\textrm{c}$)
and tensor ($V_\textrm{t}$) terms (the difference is $-5.7$~MeV for $V_\textrm{t}$
and $-7.6$~MeV for $V_\textrm{c}$).
The $^1S$ channel, which is included in $3^\textrm{(full)}_\textrm{ch}$, provides further 
decrease of the total energy ($E)$ through the central interaction; 
the decrease of $E$ is 3.8~MeV.
In contrast, in V2m-3R,  
the inclusion of  $^1S$ in $3^\textrm{(full)}_\textrm{ch}$ 
gives almost no effect. 

For the purpose of applying
AQCM-T to heavier systems such as $^8$Be, it is desirable 
to reduce the number of basis wave functions
in order to save the computational cost. For this aim, we select limited numbers of the configurations, which 
can efficiently describe the essential properties of $^4\textrm{He}$.
For G3RS2-3R, we choose four values of $k$, $k=\{1.5, 2.5, 3.5, 4.5\}$ (fm$^{-1}$) in the  
three channel calculation, which is labeled  as $3^\textrm{(4)}_\textrm{ch}$. 
As shown in Table \ref{tab:truncation},  
$3^\textrm{(4)}_\textrm{ch}$ describes 
basic properties of $^4\textrm{He}$ with almost equal quality to the full calculation, $3^\textrm{(full)}_\textrm{ch}$.
We also searched for even smaller set and found that $k=\{1.5, 2.5, 4.0\}$ (fm$^{-1}$) 
and three channels give reasonable result, labelled as
$3^\textrm{(3)}_\textrm{ch}$.
For V2m-3R, we adopt the same set as in the previous paper~\cite{AQCM-T}; 
2-channel with three $k$ values, $k=\{1.0, 2.0, 3.0\}$ (fm$^{-1}$), labeled as $2^\textrm{(3)}_\textrm{ch}$,
also gives reasonably agreement with the full calculation.

\subsection{$NN$ pair wave functions in  $^4$He}

\begin{figure*}[!htb]
\begin{center}
\includegraphics[width=16.8cm]{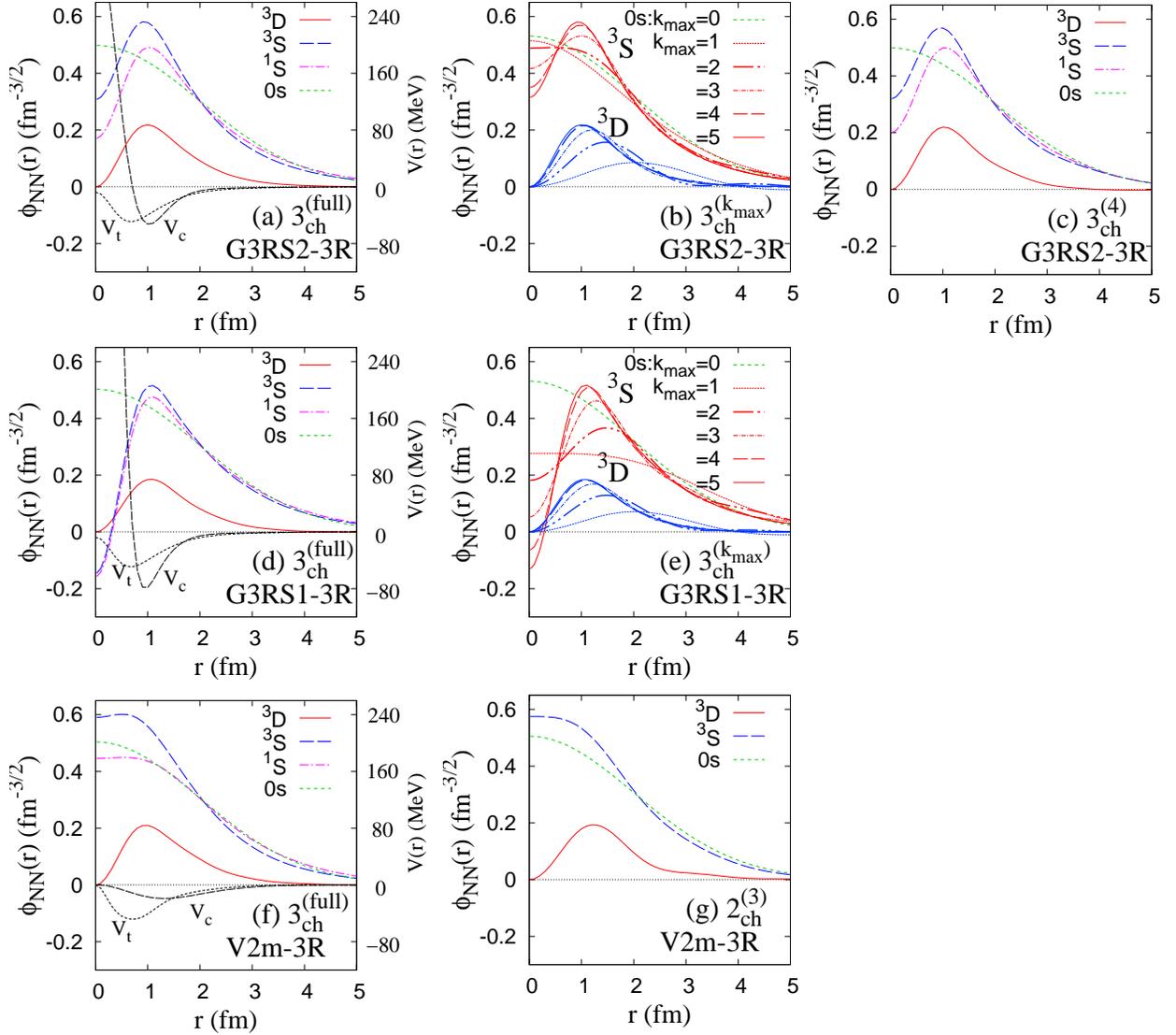} 
\end{center}
  \caption{	\label{fig:phi} (Color online) 
Pair wave functions $\phi_{NN}(r)$ 
of $^4\textrm{He}(\textrm{g.s.})$
for the $^3D$, $^3S$, and $^1S$ channels and $0s$ component 
obtained with the full configurations 
and
various interactions,
G3RS2-3R (a), G3RS1-3R (d), and V2m-3R (f).
The radial dependence of the $^3E$ parts of the central ($V_\textrm{c}$)  and tensor ($V_\textrm{t}$) interactions 
are shown in the right vertical axis. Those of
3-channel calculation with truncated $k$,
$k\le k_\textrm{max}$
obtained with
G3RS2-3R (b) and G3RS1-3R (e).
(c): $3^\textrm{(4)}_\textrm{ch}$, 3-channel calculation with $k=\{1.5,2.5,3.5,4.5\}$ fm$^{-1}$,
obtained with G3RS2-3R,
and (g): $2^\textrm{(3)}_\textrm{ch}$, 2-channel calculation with $k=\{1.0,2.0,3.0\}$ fm$^{-1}$, 
obtained with V2m-3R.
The pair wave functions of $(0s)^4$ corresponding to $k_\textrm{max}=0$ 
are shown in (b) and (e). 
}
\end{figure*}

Using the partial wave expansion  of 
$\varphi^+_k(\bvec{r})$ shown in Eq.~\eqref{eq:partial-NN},
we reconstruct the wave function of the correlated $NN$ pair
called pair wave function, noted as $\phi_{NN}(r)$.
Here, the pair wave functions 
$\phi_{NN}(r)$ are the correlated $NN$ pairs in the 
${}^{1}S$, ${}^{3}S$, and ${}^3D$ channels,  
\begin{eqnarray}  
{}^{1}S:&&\phi_g(\bvec{r}_{g})\phi_g(\bvec{r}'_{g})\otimes \nonumber\\
&&\phi^{^1S}_{NN}(r)\varphi^{(0)}_{0}(r')  \otimes Y_{00} Y_{00}  \otimes \chi^\sigma_{0} \chi^\sigma_{0}
\otimes [ \chi^\tau_{1} \chi^\tau_{1}]_{T=0},\nonumber
\end{eqnarray}
\begin{eqnarray}  
{}^{3}S:&&\phi_g(\bvec{r}_{g})\phi_g(\bvec{r}'_{g})\otimes \nonumber\\
&&\phi^{^3S}_{NN}(r)\varphi^{(0)}_{0}(r') \otimes Y_{00} Y_{00} \otimes [\chi^\sigma_{1} \chi^\sigma_{1}]_{S=0}\otimes
\chi^\tau_{0} \chi^\tau_{0}, \nonumber
\end{eqnarray}
and 
\begin{eqnarray}  
{}^{3}D:&&\phi_g(\bvec{r}_{g})\phi_g(\bvec{r}'_{g})\otimes\nonumber \\
&&\phi^{^3D}_{NN}(r)\varphi^{(0)}_{0}(r') \otimes \left[ Y_{20} Y_{00}  \otimes 
 [\chi^\sigma_{1} \chi^\sigma_{1}]_{S=2}\right]_{J=0}\otimes
\chi^\tau_{0} \chi^\tau_{0},\nonumber
\end{eqnarray}
 respectively.

The pair wave functions in ${}^4\textrm{He}$ are shown in Fig.~\ref{fig:phi}. 
The radial $(r)$ dependence of the $^3E$ parts of the central ($V_c$)  and tensor ($V_t$) interactions 
are also shown to see the correspondence.

Figure~\ref{fig:phi} (a) shows the pair wave functions of the full calculation ($3_\textrm{ch}^\textrm{(full)}$)
obtained with G3RS2-3R. 
Here
$\phi^{^3D}_{NN}(r)$
(line ``$^3D$")
shows a peak in the middle-range region, $r\sim 1$ fm.
Also,
$\phi^{^3S}_{NN}(r)$ and $\phi^{^1S}_{NN}(r)$ 
(lines ``$^3S$" and ``$^1S$")
are peaked
in the middle-range ($r\sim 1$~fm) region corresponding to the 
pocket of the central-even interaction,
but they are strongly suppressed in the $r< 1$~fm  region
because of the short-range repulsion.
In the middle-range region,
the amplitude of $\phi^{^3S}_{NN}(r)$ 
is larger than
$\phi^{^1S}_{NN}(r)$ because of the additional attraction
caused by the tensor interaction
incorporated through the $^3S$-$^3D$ coupling.
Since this additional attraction is particularly strong at the peak position of 
$\phi^{^3D}_{NN}(r)$, it enhances the amplitude of $\phi^{^3S}_{NN}(r)$ in this $r\sim 1$~fm region and efficiently contributes to the energy gain of the 
central-even interaction. 
This appearance of distinct peaks in the $^1S$ and $^3S$ channels
is a unique feature of the realistic interactions, 
G3RS2-3R (Fig.~\ref{fig:phi} (a))
and 
G3RS1-3R (Fig.~\ref{fig:phi} (d)), 
but cannot be 
seen in the case of effective interactions without the short-range core of the central interaction as in 
V2m-3R (Fig.~\ref{fig:phi} (f)).

Figure~\ref{fig:phi} (b) is also for G3RS2-3R, 
but this shows the result of $3^{k\le k_\textrm{max}}_\textrm{ch}$
with different $k_\textrm{max}$ values. 
As $k_\textrm{max}$ increases, the amplitude of $\phi^{^3D}_{NN}(r)$ 
(lines ``$^3D$")
rapidly grows, 
where the configurations of $k\lesssim 3$   fm$^{-1}$ largely contribute 
to the peak in the middle-range region ($r\sim1$~fm).
For $\phi^{^3S}_{NN}(r)$ (``$^3S$"),
the middle-range part grows up and the short-range part is suppressed 
with increasing $k_\textrm{max}$.
In particular, 
the high-momentum configurations with $k=3\sim 4$~fm$^{-1}$ contribute to both the enhancement of the
middle-range part and the suppression of the short-range part,
whereas the configurations with lower momenta ($k\lesssim 2$~fm$^{-1}$) 
mainly contribute to the reduction of the tail part in the long-range region around $r\sim 3$~fm compared with the 
uncorrelated $0s$ state (line ``$0s$:$k_\textrm{max}=0$").

In Fig.~\ref{fig:phi}~(c),
the pair wave functions for G3RS2-3R 
obtained with reduced number of the basis states, $3_\textrm{ch}^{(4)}$, 
are shown, which
reasonably reproduce the basic properties  
of the full 3-channel calculation, $3_\textrm{ch}^\textrm{(full)}$. 

Figures~\ref{fig:phi} (d) and (e) show the pair wave functions of G3RS1-3R, wheres (f) and (g)  are for V2m-3R.
For G3RS1-3R, 
$\phi^{^3S}_{NN}(r)$ and $\phi^{^1S}_{NN}(r)$
(lines ``$^3S$" and ``$^1S$" in Fig.~\ref{fig:phi} (d))
show
further suppression of the amplitudes in the short range and enhancement of a narrower peak in the middle range
compared with G3RS2-3R
(lines ``$^3S$" and ``$^1S$" in Fig.~\ref{fig:phi} (a)),
because of higher repulsion (in short range) 
and deeper attraction (in the middle range)
of the central-even interaction. 
For V2m-3R, 
the $S$-wave pair wave functions, $\phi^{^3S}_{NN}(r)$ and $\phi^{^1S}_{NN}(r)$
(lines ``$^3S$" and ``$^1S$" in Fig.~\ref{fig:phi} (f)),
do not show such suppression in the short range nor remarkable peak 
in the middle range}
since V2m-3R has no repulsive core. 
Thus, $S$-wave pair wave functions, 
$\phi^{^3S}_{NN}(r)$ and $\phi^{^1S}_{NN}(r)$, show different features
in the cases of G3RS2-3R and V2m-3R
depending on the presence of  
 the repulsive core of the central interaction,
even though the contributions of the central interactions are similar in the calculated energy.
On the other hand, the $D$-wave pair wave function ($\phi^{^3S}_{NN}(r)$) shows quite similar features
in the three cases of G3RS2-3R, G3RS1-3R, and V2m-3R.

\section{Application to $^8\textrm{Be}$} \label{sec:results-be8}

\begin{table*}[!tbhp]
\caption{Each component of the Hamiltonian 
and the probabilities of the $0s$ and $^3D$ components in
$^8$Be obtained with the fixed-$d$  calculation for G3RS2-3R and V2m-3R. 
For energies, the expectation values of the total energy $(E)$, kinetic energy $(T)$, central interaction $(V_\textrm{c})$, and tensor interaction $(V_\textrm{t})$ are shown. 
The values for the ideal $^4\textrm{He}$(g.s.)-$^4\textrm{He}$(g.s.) state
in the asymptotic region of $d_\alpha\to \infty$ and 
threshold ($2\alpha$ thres.) energies are evaluated 
as twice the $^4\textrm{He}$ energy obtained with a consistent model space.
The constant shift of $T_r=\hbar\omega/4=5.2$ MeV is added
to the kinetic and total energies for the ideal $^4\textrm{He}$(g.s.)-$^4\textrm{He}$(g.s.) state,
corresponding to the localization of clusters with fixed relative distance.
The values of the $0s$ state for the conventional two-$\alpha$ cluster model 
with the V2 interaction are also shown. All energies are in MeV.
\label{tab:be8}
}
\begin{center}
\begin{tabular}{crrrrrrrrrrrrrrrrrrrrr}
\hline
\hline
\multicolumn{7}{c}{G3RS2-3R}	\\													
$d$	(fm) &$	\langle E\rangle_{2\alpha}	$&$	\langle T\rangle_{2\alpha}	$&$	\langle V_\textrm{c}\rangle_{2\alpha}	$&$	\langle V_\textrm{t}\rangle_{2\alpha}	$&$	{\cal P}_{0s}	$&$	{\cal P}_{^3D}	$\\
1 	&$	24.8 	$&$	153.1 	$&$	-114.5 	$&$	-17.8 	$&$	0.95 	$&$	0.04 	$\\
2 	&$	-2.7 	$&$	157.8 	$&$	-121.0 	$&$	-43.8 	$&$	0.89 	$&$	0.09 	$\\
3 	&$	-29.2 	$&$	153.8 	$&$	-122.3 	$&$	-65.0 	$&$	0.85 	$&$	0.11 	$\\
4 	&$	-39.0 	$&$	144.5 	$&$	-117.6 	$&$	-69.9 	$&$	0.84 	$&$	0.12 	$\\
5 	&$	-40.5 	$&$	140.1 	$&$	-114.4 	$&$	-69.9 	$&$	0.84 	$&$	0.12 	$\\
6 	&$	-40.6 	$&$	139.3 	$&$	-113.6 	$&$	-69.8 	$&$	0.84 	$&$	0.12 	$\\
GCM	&$	-45.8 	$&$	138.2 	$&$	-116.8 	$&$	-70.9 	$&$	0.83 	$&$	0.12 	$\\
$2\alpha$ thres.	&$	-50.3 	$&$		$&$		$&$		$&$		$&$		$\\
	&$	2\langle E\rangle_{\alpha}+T_r	$&$	2\langle T\rangle_{\alpha}+T_r	$&$	2\langle V_\textrm{c}\rangle_{\alpha}	$&$	2\langle V_\textrm{t}\rangle_{\alpha}	$&$	\{ {\cal P}_{0s}\}^2_\alpha	$&$		$\\
$^4\textrm{He}$(g.s.)-$^4\textrm{He}$(g.s.)	&$	-45.1 	$&$	151.0 	$&$	-116.9 	$&$	-82.1 	$&$	0.78	$&$		$\\
\hline
\multicolumn{7}{c}{V2m-3R}	\\
$d$	(fm) &$	\langle E\rangle_{2\alpha}	$&$	\langle T\rangle_{2\alpha}	$&$	\langle V_\textrm{c}\rangle_{2\alpha}	$&$	\langle V_\textrm{t}\rangle_{2\alpha}	$&$	{\cal P}_{0s}	$&$	{\cal P}_{^3D}	$\\
1 	&$	3.4 	$&$	148.1 	$&$	-132.8 	$&$	-16.1 	$&$	0.94 	$&$	0.04 	$\\
2 	&$	-18.6 	$&$	149.1 	$&$	-132.3 	$&$	-40.0 	$&$	0.90 	$&$	0.09 	$\\
3 	&$	-39.7 	$&$	142.3 	$&$	-127.0 	$&$	-59.8 	$&$	0.86 	$&$	0.12 	$\\
4 	&$	-47.4 	$&$	132.3 	$&$	-119.7 	$&$	-64.4 	$&$	0.86 	$&$	0.12 	$\\
5 	&$	-47.8 	$&$	127.6 	$&$	-115.3 	$&$	-64.2 	$&$	0.86 	$&$	0.12 	$\\
6 	&$	-47.4 	$&$	126.7 	$&$	-113.9 	$&$	-64.1 	$&$	0.86 	$&$	0.12 	$\\
GCM	&$	-51.3 	$&$	124.2 	$&$	-115.4 	$&$	-64.2 	$&$	0.82 	$&$	0.12 	$\\
$2\alpha$ thres.	&$	-56.4 	$&$		$&$		$&$		$&$		$&$		$\\
	&$	2\langle E\rangle_{\alpha}+T_r	$&$	2\langle T\rangle_{\alpha}+T_r	$&$	2\langle V_\textrm{c}\rangle_{\alpha}	$&$	2\langle V_\textrm{t}\rangle_{\alpha}	$&$	\{ {\cal P}_{0s}\}^2_\alpha	$&$		$\\
$^4\textrm{He}$(g.s.)-$^4\textrm{He}$(g.s.)	&$	-51.2 	$&$	134.9 	$&$	-114.9 	$&$	-74.5 	$&$	0.78	$&$		$\\
\hline
\multicolumn{7}{c}{V2:0s}	\\												
$d$	(fm) &$	\langle E\rangle_{2\alpha}	$&$	\langle T\rangle_{2\alpha}	$&$	\langle V_\textrm{c}\rangle_{2\alpha}	$&$	\langle V_\textrm{t}\rangle_{2\alpha}	$&$	{\cal P}_{0s}	$&$	{\cal P}_{^3D}	$\\
1 	&$	-38.5 	$&$	145.3 	$&$	-187.6 	$&$	0 	$&$	1 	$&$	0 	$\\
2 	&$	-45.7 	$&$	132.1 	$&$	-181.4 	$&$	0 	$&$	1 	$&$	0 	$\\
3 	&$	-51.8 	$&$	116.0 	$&$	-171.1 	$&$	0 	$&$	1 	$&$	0 	$\\
4 	&$	-53.0 	$&$	104.1 	$&$	-160.2 	$&$	0 	$&$	1 	$&$	0 	$\\
5 	&$	-51.3 	$&$	99.5 	$&$	-153.5 	$&$	0 	$&$	1 	$&$	0 	$\\
6 	&$	-50.2 	$&$	98.6 	$&$	-151.3 	$&$	0 	$&$	1 	$&$	0 	$\\
GCM	&$	-55.5 	$&$	100.7 	$&$	-159.2 	$&$	0 	$&$	 	$&$	0 	$\\
$2\alpha$ thres.&$	-55.8 	$&$		$&$		$&$		$&$		$&$		$\\
	&$	2\langle E\rangle_{\alpha}+T_r	$&$	2\langle T\rangle_{\alpha}+T_r	$&$	2\langle V_\textrm{c}\rangle_{\alpha}	$&$	2\langle V_\textrm{t}\rangle_{\alpha}	$&$	\{ {\cal P}_{0s}\}^2_\alpha	$&$		$\\
$^4\textrm{He}$(g.s.)-$^4\textrm{He}$(g.s.)	&$	-50.6 	$&$	98.5 	$&$	-150.7 	$&$	0 	$&$	1 	$&$	0 	$\\
\hline
\hline
\end{tabular}
\end{center}
\end{table*}

\begin{figure*}[!tbhp]
\begin{center}
\includegraphics[width=15cm]{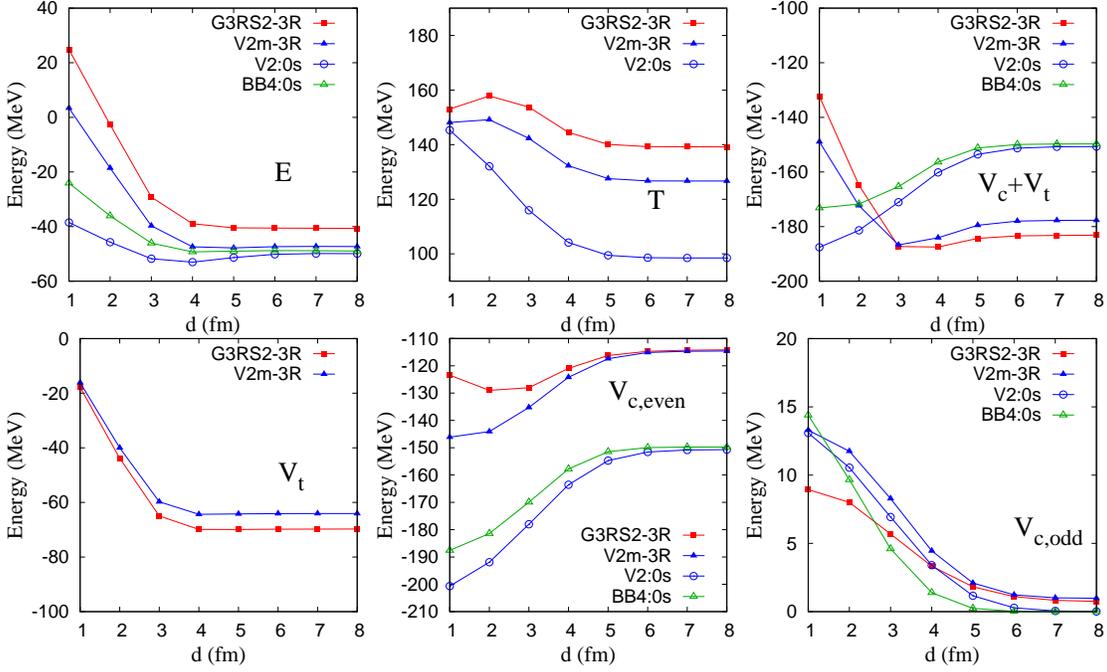} 
\end{center}
  \caption{	\label{fig:ene}  (Color online) 
Each component of the Hamiltonian for
$^8$Be 
obtained with full configurations ($3^\textrm{(full)}_\textrm{ch}$) 
but fixed $d$.
The results of G3RS2-3R and V2m-3R 
are compared, and the
values of the $0s$ state for the V2 
(V2:$0s$)
and BB4 
(BB4:$0s$)
interactions are also shown.
The expectation values of the total energy ($E$), 
kinetic energy ($T$), 
sum of central and tensor interactions
($V_\textrm{c}$+$V_\textrm{t}$)
are shown in the top-left, top-middle, and top-right panels, and those 
of the tensor ($V_\textrm{t}$), 
central-even ($V_\textrm{c,even}$), 
and central-odd ($V_\textrm{c,odd}$)
components are shown in 
the bottom-left, bottom-middle, and bottom-right panels, respectively.
}
\end{figure*}

\begin{figure}[!thbp]
\begin{center}
\includegraphics[width=6cm]{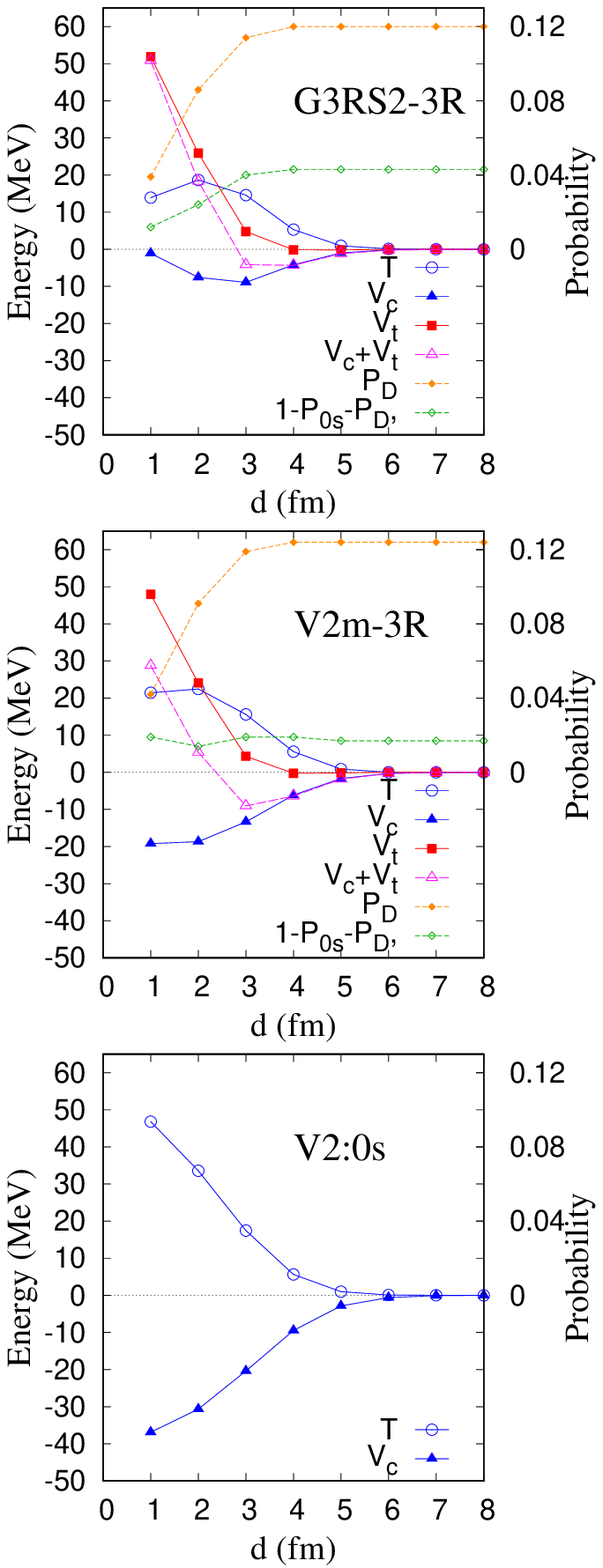} 
\end{center}
  \caption{\label{fig:be8-offset} (Color online) 
 Each component of the Hamiltonian 
 (left vertical axis)
 and $D$ state probability 
 (right vertical axis)
 of
 $^8$Be obtained with the full configurations ($3^\textrm{(full)}_\textrm{ch}$)
of AQCM-T but with fixed $d$.
In each panel, the expectation values of the kinetic energy ($T$), the central ($V_\textrm{c}$) and 
the tensor ($V_\textrm{t}$) interactions, and sum of central and tensor interactions
($V_\textrm{c}+V_\textrm{t}$) are shown.
Relative energies are measured from the values at $d_\textrm{max}$ $(=8~\text{fm})$
are plotted as functions of $d$. 
The results for G3RS2-3R, V2m-3R,
and V2:$0s$ are shown in the 
top, middle, and bottom panels, respectively.
Right vertical axis: The probability of the $^3D$ component  (${\cal P}_{^3D}$) and that of the correlated $S$-wave component
 ($1-{\cal P}_{0s}-{\cal P}_{^3D}$).}
\end{figure}

In this section, we investigate  $^8$Be with the present AQCM-T method.
To reduce the calculation costs, 
the correlated $\alpha$ cluster of $3^{(4)}_\textrm{ch}$ 
is adopted for G3RS2-3R, whereas
$2^{(3)}_\textrm{ch}$ is adopted for V2m-3R, which can reasonably 
describe basic properties of $^4\textrm{He}$. These are fixed  
in all the results of $^8$Be in this section. 
They are compared with V2:$0s$, the conventional two-$\alpha$ cluster
calculation with the V2 interaction.

\subsection{Fixed-$d$ calculation of $^8\textrm{Be}$}
The energies and probabilities of $^3D$ and $0s$ components obtained with 
the fixed-$d$ and full GCM calculations 
are compared in Table \ref{tab:be8}.
We also show the 
values for the ideal $^4\textrm{He}$(g.s.)-$^4\textrm{He}$(g.s.) state
in the asymptotic region of $d_\alpha\to \infty$, which 
are evaluated as twice the $^4\textrm{He}$ energies calculated with the consistent model space.
Note that 
a constant energy shift of
$T_r=\langle \widehat{T}_\mathrm{G} \rangle/3=\hbar\omega/4$ 
is added to the kinetic and total energies  for the ideal $^4\textrm{He}$(g.s.)-$^4\textrm{He}$(g.s.) state,
which corresponds to the increase of the energy due to the
localization of the inter-cluster motion by fixing $d$, even though $d$ is quite large.
The two-$\alpha$ threshold energy estimated from  
$^4\textrm{He}(\textrm{g.s.})$ is also shown.

As the distance $d$ increases,
each energy component gets closer to 
the values for the 
ideal $^4\textrm{He}$(g.s.)-$^4\textrm{He}$(g.s.) state,
indicating that two 
$^4\textrm{He}$ come to their ground state.
However, even at large $d$ values, 
there still remain some 
differences compared with the  values for the ideal 
$^4\textrm{He}(\textrm{g.s.})$-$^4\textrm{He}(\textrm{g.s.})$ state;
the total energy 
($\langle E\rangle_{2\alpha}$)
is higher by about 4 MeV, and the absolute values of 
kinetic 
($\langle T\rangle_{2\alpha}$)
and tensor 
$\langle V_\textrm{t}\rangle_{2\alpha}$)
energies are smaller by about 10~MeV.
Also, the $0s$ probability 
(${\cal P}_{0s}$)
is smaller by about 5\%.
These differences may originate from the limitation of the 
present AQCM-T framework for $^8\textrm{Be}$;
the second-order correlation effect
that both of two $\alpha$ clusters are simultaneously 
correlated, is omitted. 
From the energy at $d_\textrm{max}$ $(=8\ \text{fm})$,
this defect of the binding energy due to the missing second-order correlation
is estimated to be 5.4 MeV and 3.1 MeV in  G3RS2-3R and V2m-3R, respectively.
This artifact should be taken into account in the following discussion.

As two $\alpha$ clusters approach, the total energy increases because of the 
increase of the kinetic energy and decrease of the 
attractive effect of the
potential energy.
In particular, the reduction
of the tensor attraction significantly contributes
in the $d\le 2$~fm region, in which the $D$-state probability is strongly suppressed.
In~Figs.~\ref{fig:ene} and \ref{fig:be8-offset},
the energy and probability of each channel are plotted as functions of $d$.
The former is for
the absolute energies, whereas the latter is for the relative energies measured from $d_\textrm{max}$ $(=8~\textrm{fm})$.
The ${\cal P}_{^3D}$ probability and that for the 
correlated $S$-wave pair orthogonal to the uncorrelated $(0s)^4$ state for each $\alpha$
 ($1-{\cal P}_{0s}-{\cal P}_{D}$)
are also plotted.
In the case of V2:$0s$ (open circle in Fig.~\ref{fig:ene}), 
the total energy ($E$) somewhat increases 
when two $\alpha$ clusters get closer
mainly 
because of the increases in the kinetic ($T$)
energy and central-odd interaction ($V_\textrm{c,odd}$).
The former is the Pauli blocking effect between $\alpha$ clusters;
four nucleons in one of the $\alpha$ clusters are raised up to 
the $0p$-orbits from the $0s$-orbits, which
contributes in the $d \le 3$~fm region.
The latter is responsible for the repulsion in
the $d \le 4$~fm region ($d \le 3$~fm in the BB4:$0s$ (open triangle) case).
These two repulsions are long-distance effects, which 
are rather general features of the $\alpha$-$\alpha$ potential 
and seen both in G3RS2-3R (solid square)
and 
V2m-3R (solid triangle).
It should be commented that the effect coming from the central-odd part 
 in G3RS2-3R is smaller
by a factor of 2/3 than in V2m-3R and V2:$0s$.

However, in the G3RS2-3R 
(solid square in Fig.~\ref{fig:ene})
and 
V2m-3R 
(solid triangle)
cases, another repulsive effect at short distances comes from the tensor suppression.
In the long-distance region ($d\sim 8$ fm), 
the tensor interaction 
($V_\textrm{t}$)
significantly 
contributes to the binding energy of each $\alpha$ cluster 
through the mixing of the $^3D$ state.
Although the $^3D$-state mixing induces some increase of the kinetic energy
($T$), this is 
compensated by the larger gain of the attractive tensor contribution. 
As $\alpha$ clusters come close to each other,
this tensor contribution is reduced, and
the $^3D$ state {
(${\cal P}_{D}$ in Fig.~\ref{fig:be8-offset})
is suppressed because of the Pauli blocking of nucleons in two $\alpha$ clusters.
These effects works
in the region of $d\le 2$~fm,
relatively shorter distance region compared with the repulsive effects of the kinetic and central-odd parts 
in the longer distance region. 
Note that weakening of the 
kinetic energy in this region is also a signal of the tensor suppression.
Comparing the results of G3RS2-3R and V2m-3R, the repulsive effect at short $d$
is slightly larger in G3RS2-3R  than in V2m-3R; the energy is higher by about 15~MeV at $d=1$~fm. 
This is attributed to the additional effect of the central-even interaction (mainly $^3E$)
caused by the reduction of the correlated $S$-wave pair, $1-{\cal P}_{0s}-{\cal P}_{D}$,
which can be seen in the $d\le 2$~fm region in the case of G3RS2-3R (Fig.~\ref{fig:be8-offset} (a)),
but not in V2m-3R (Fig.~\ref{fig:be8-offset} (b)).
These repulsive effects at short distances related to the tensor 
($V_\textrm{t}$ in Fig.~\ref{fig:ene})
and central-even interactions 
($V_\textrm{c,even}$)
are attributed to
the high-momentum correlations of nucleons in the 
$\alpha$ clusters.
More detailed discussions are given later in subsection \ref{sec:discussion-be8}.

\subsection{Full GCM calculation for $^8\textrm{Be}$}

\begin{figure}[!tbhp]
\begin{center}
\includegraphics[width=6.5cm]{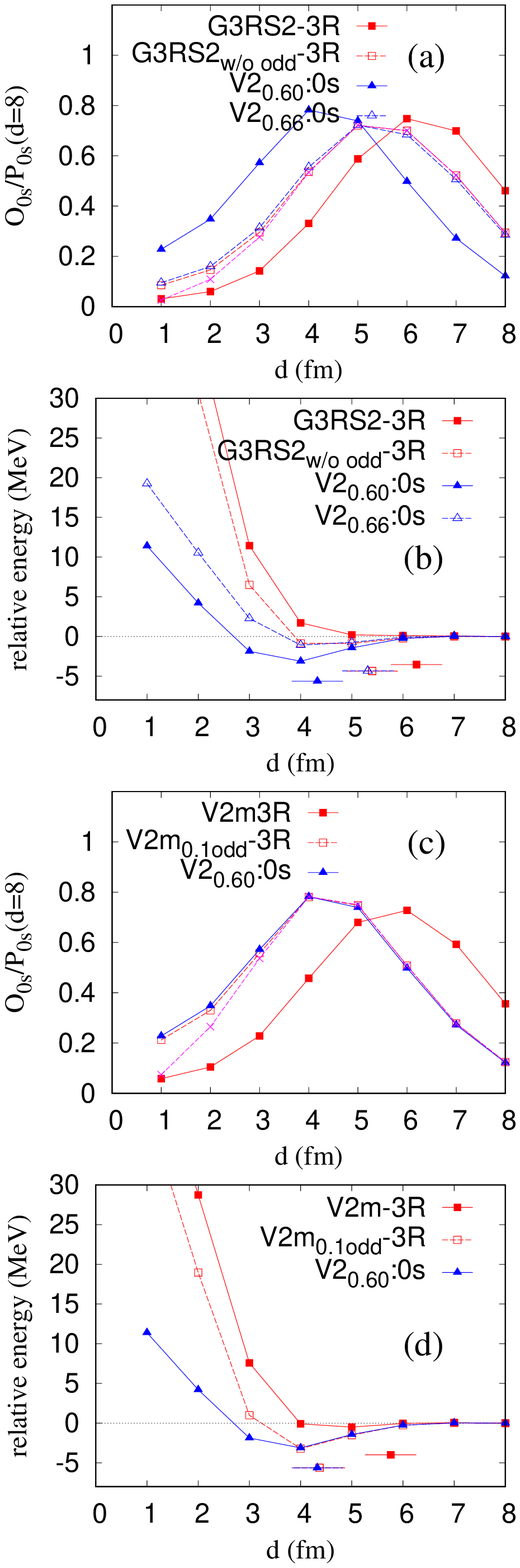} 
\end{center}
  \caption{	\label{fig:be8-pa}  (Color online) 
Squared overlap at $d$ between the full 
GCM calculation of AQCM-T and the $0s$ state of $^8\textrm{Be}$.
(a): G3RS2-3R and that without the odd part of the central interaction 
(G3RS$_{\rm w/o\ odd}$)
are compared with V2:0s with the Majorana parameter of $m=0.60$ (V2$_{0.60}$:0$s$)
and $0.66$ (V2$_{0.66}$:0$s$).
(c): V2m-3R with the default value for the central-odd part
and weakened one with a factor of 0.1 (V2m$_{\rm 0.1odd}$-3R)
are compared with V2:$0s$ with $m=0.60$ (V2$_{0.60}$:0$s$).
The squared overlap is scaled by $1/{\cal P}_{0s}$ with the value of ${\cal P}_{0s}$ at $d_\textrm{max}$ $(=8\ \text{fm})$.
Squared overlap with the frozen-$\alpha$ wave function is also shown 
as cross symbols in (a) and (c).
Relative energy measured from $d_\textrm{max}$ $(=8~\text{fm})$ with
fixed-$d$ calculations for (b): G3RS2-3R and (d): V2m-3R. 
}
\end{figure}

We perform full GCM calculation by superposing the wave functions with different values for the generator coordinate $d$ ($d=1,\ldots, 8$~fm). 
The calculated energy and probability  
for G3RS2-3R, V2m-3R, and V2:$0s$ are shown in the row ``GCM'' of
Table~\ref{tab:be8}.
Compared with the minimum energy of the fixed-$d$ calculation,  
the total energy is lower by about 5~MeV
because the kinetic energy coming from the localization 
of the inter-cluster motion is released by the superposition of different $d$ configurations in the GCM calculation. 
The relative energy $E_r$ of $^8\textrm{Be}(0^+)$ measured from the two-$\alpha$ threshold is 7.8~MeV, 5.0~MeV, and
0.3~MeV for  G3RS2-3R, V2m-3R, and V2:$0s$, respectively.  The experimental value of $E_r=0.1$~MeV is well reproduced with V2:$0s$, but 
not with G3RS2-3R and V2m-3R.   
This shortcoming dominantly comes from the missing second-order correlation 
in the asymptotic region. As mentioned previously, this effect is 5.4~MeV (3.1~MeV) for G3RS2-3R (V2m-3R). However, even with this effect, the energy is still higher by a few MeV,
implying that some attractive effect is still missing.  
We do not know its reason,  but three-body interactions might help for the additional attraction.

The solid squares in
Fig.~\ref{fig:be8-pa}~(a) and (c) 
show the squared overlap
between the full GCM solution and the $0s$ state,
$|\langle \Phi^{0s}_{2\alpha,0^+}(d)|\Psi^\textrm{GCM}_{2\alpha,0^+}\rangle|^2$,
for G3RS2-3R and V2m-3R, respectively,
which are
normalized to the $0s$ probability (${\cal P}_{0s}$) obtained 
at $d_\textrm{max}=8$~fm. 
The total energies of the fixed-$d$ calculation are shown 
as solid squares
in Fig.~\ref{fig:be8-pa}~(b) and (d) for G3RS2-3R and V2m-3R, respectively, which show
the relative energies measured from the value at $d_\textrm{max}=8$~fm. 
Here the energies of the full GCM results are also shown as the horizontal lines.
Compared with V2:$0s$, both amplitudes of G3RS2-3R and V2m-3R
are pushed out 
because of the less two-$\alpha$ binding. 

In order to remove this artifact coming from the less binding, 
we tune the odd part of the central interaction of G3RS2-3R and V2m-3R so that 
the GCM calculations give the same two-$\alpha$ bindings 
as the V2:$0s$ calculation.
In Fig.~\ref{fig:be8-pa}, we adjust the relative energies measured from $d_\textrm{max}$ $(=8~\text{fm})$ 
obtained with the
fixed-$d$ calculations.
For V2m-3R, we reduce the odd part of the central interaction to 
10\% of the original strength 
(V2m$_{\rm 0.1odd}$-3R)
and compare with V2:$0s$ with 
the default Majorana parameter of $m=0.60$ (V2$_{0.60}$:0$s$).
For G3RS2-3R, we remove the odd part of the central interaction 
(G3RS$_{\rm w/o\ odd}$)
and compare with
V2:$0s$ with $m=0.66$
(V2$_{0.66}$:0$s$),
corresponding to the enhancement of the central-odd interaction by 80\%.
The $0s$ normalized overlap for G3RS2-3R 
(Fig.~\ref{fig:be8-pa}~(a) solid square)
and V2m-3R (Fig.~\ref{fig:be8-pa}~(c) solid squre) 
show similar $d$-dependence of to
that of the V2:$0s$ results (solid triangle), which is
a little bit surprising because  
the $\alpha$-$\alpha$ energy curves of G3RS2-3R 
(Fig.~\ref{fig:be8-pa}~(b) solid square) 
and V2m-3R 
(Fig.~\ref{fig:be8-pa}~(d) solid square)
show the repulsion at short distances
compared with V2:$0s$ (solid triangle). 

It should be commented that 
the normalized $0s$ overlap introduced here is a good measure of  
the two-$\alpha$ structure at $d$.
However, this quantity may contain not only the unpolarized two-$\alpha$ component 
but also the polarized component. 
To extract only the unpolarized $\alpha$ component,  
we calculate the squared overlap between $\Psi^\textrm{GCM}_{2\alpha,0^+}$ and 
the frozen-$\alpha$ wave function $\Psi^\textrm{frozen}_{2\alpha,0^+}(d)$ defined in Eq.~\eqref{eq:frozen-cluster-8Be}.
The result 
is shown as cross symbols in Fig.~\ref{fig:be8-pa} (a) and (c).
In the $d \ge 3$~fm region,  there is no difference,
because the core polarization is small in this region. 
At shorter distances, $d\le 2$~fm, 
the internal structure of the $\alpha$ clusters changes mainly because of the reduction of the 
$^3D$ component, 
and the overlap with the frozen-$\alpha$ calculation shows 
the suppression of the two-$\alpha$ probability compared with the normalized $0s$ overlap.

\subsection{Discussions for the two-$\alpha$ system} \label{sec:discussion-be8}

\begin{figure*}[!tbhp]
\begin{center}
\includegraphics[width=16.8cm]{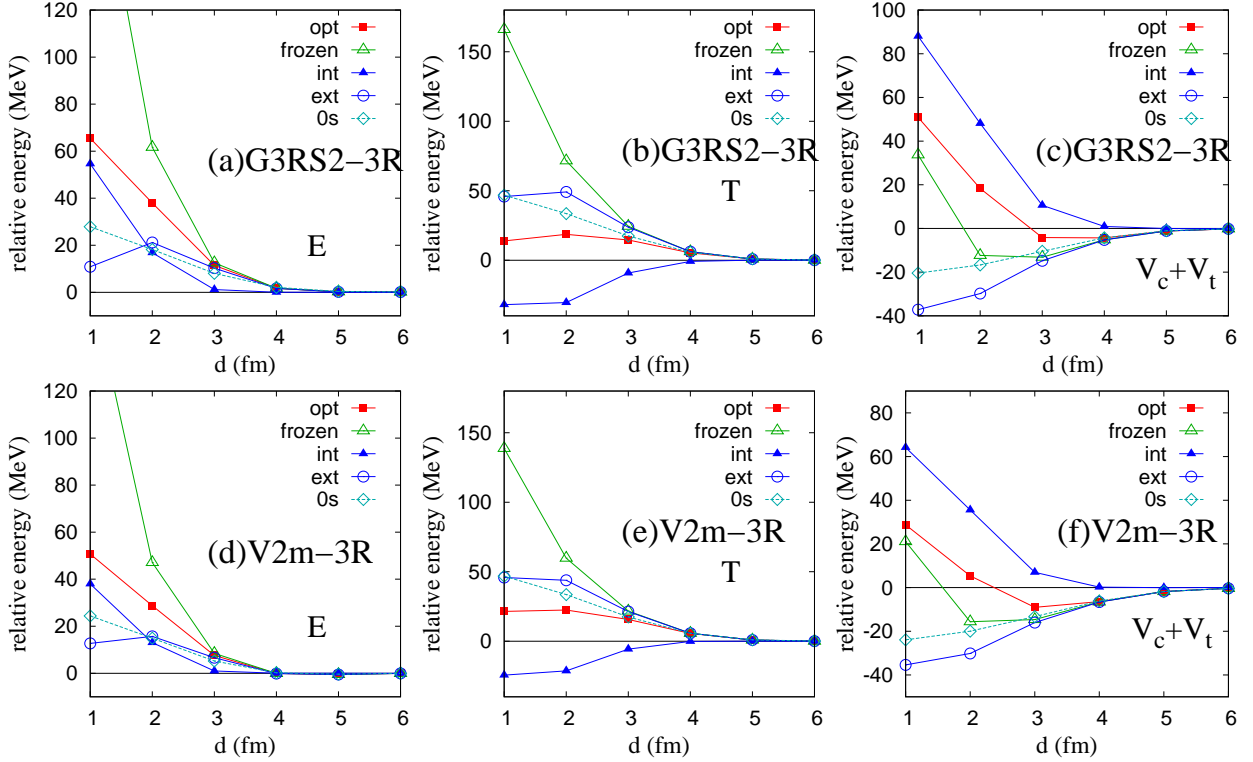} 
\end{center}
  \caption{ \label{fig:ene-g3rs2-v2m-1} (Color online) 
Energy for each component of the Hamiltonian of
$^8$Be obtained by the optimized-$\alpha$ (fixed-$d$) calculation (opt) and 
frozen-$\alpha$ calculation (frozen), and the conventional 
two-$\alpha$
cluster model with the $0s$ configuration ($0s$).  
The internal energy of $\alpha$ clusters (int)
and external one between them (ext) are also shown. 
Upper panels: expectation values of the total energy 
($E$, left), 
kinetic energy ($T$, middle), 
 and the sum of the central and tensor interactions
 ($V_\textrm{c}+V_\textrm{t}$, right) 
 for G3RS2-3R.  Lower panels: those for V2m-3R.
Relative energies measured from the values at $d_\textrm{max}~(=8~\text{fm})$
are plotted as functions of $d$. 
}
\end{figure*}

\begin{figure*}[!tbhp]
\begin{center}
\includegraphics[width=16.8cm]{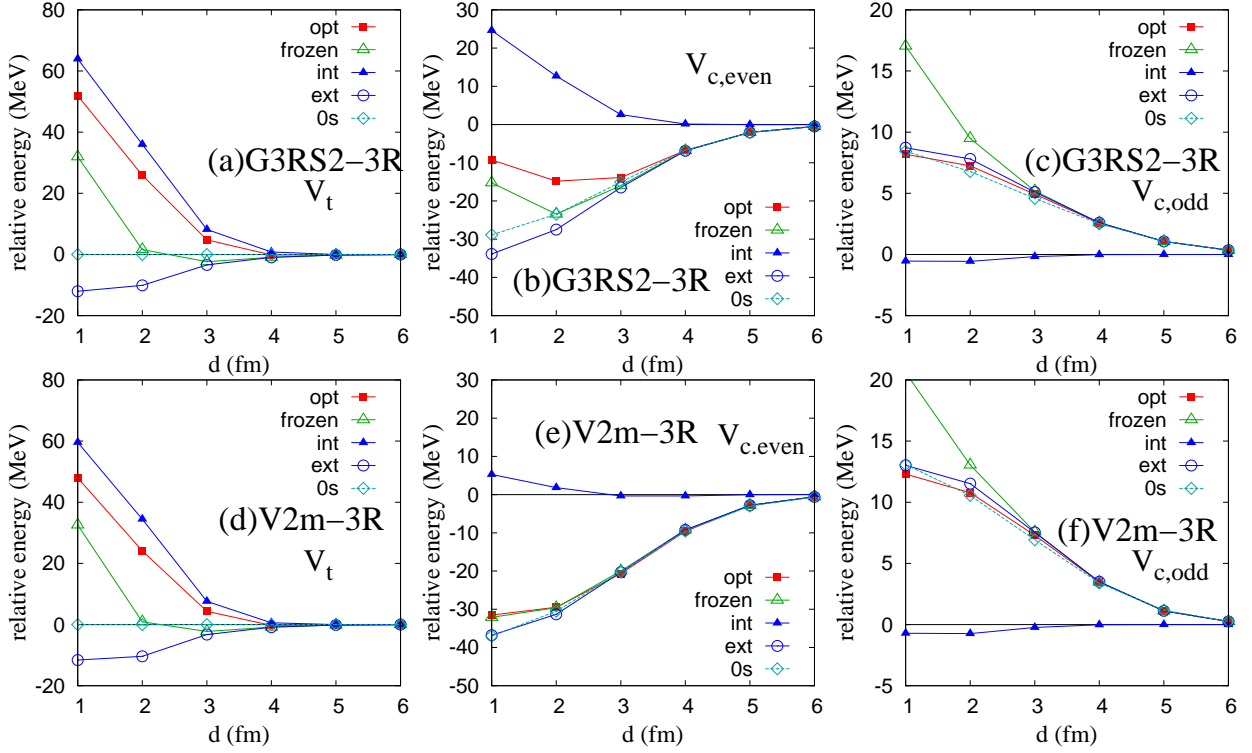} 
\end{center}
  \caption{	\label{fig:ene-g3rs2-v2m-2} (Color online) 
Same as Fig.~\ref{fig:ene-g3rs2-v2m-1} but for the tensor 
($V_\textrm{t}$,
left), central-even 
($V_\textrm{c.even}$,
middle), and central-odd
($V_\textrm{c.odd}$,
 right) interactions.}
\end{figure*}

\begin{figure*}[!tbhp]
\begin{center}
\includegraphics[width=16.8cm]{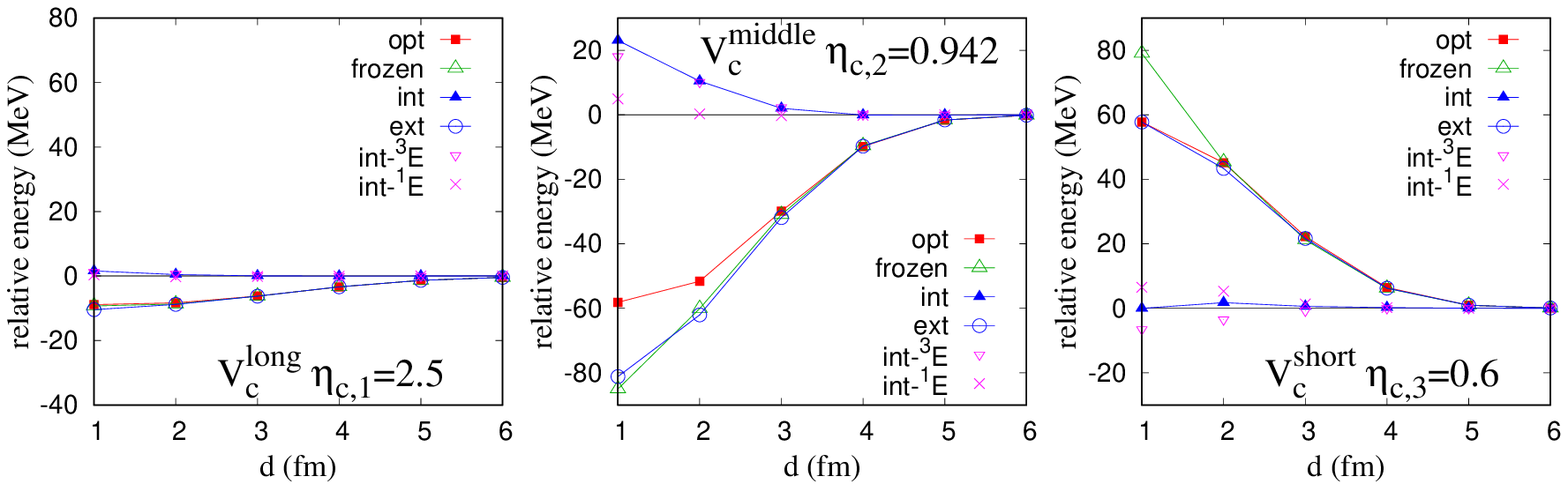} 
\end{center}
  \caption{\label{fig:ene-g3rs2-v2m-3} (Color online) 
Same as Fig.~\ref{fig:ene-g3rs2-v2m-1} but for 
the long-range ($\eta_{c,1}=$2.5 fm), middle-range ($\eta_{c,2}=$0.942 fm), and 
short-range ($\eta_{c,3}=$0.6 fm) terms for the central-even part of
G3RS2-3R are shown in the left, middle, and right, respectively. 
For the contribution of the internal energy, 
$^3E$ 
(int-$^3E$)
and $^1E$ 
(int-$^1E$)
parts 
are shown in addition to their sum (int).
}
\end{figure*}

In order to clarify the roles of the tensor and short-range correlations
in the two-$\alpha$ system, here we discuss the $d$ dependence of 
the energy in more detail. 
We focus on the origin of the repulsive effects in the 
short-distance region ($d\le 2$ fm) of the two-$\alpha$ system.

In Figs.~\ref{fig:ene-g3rs2-v2m-1}, \ref{fig:ene-g3rs2-v2m-2}, and \ref{fig:ene-g3rs2-v2m-3}, 
we show the total energy and the contribution of each term of the Hamiltonian obtained by the 
and frozen-$\alpha$ calculations as well as those for the $0s$ state
($(0s)^4$ configuration for each $\alpha$ cluster).
In order to stress 
the optimization of the correlated $\alpha$ cluster at each $d$,  
we call here the fixed-$d$ calculation ``optimized-$\alpha$ calculation''. 
We also show the contribution of the internal excitation energy 
of the $\alpha$ clusters
caused by the changes in the internal structure 
and that of the external energy between $\alpha$ clusters
defined in subsection \ref{subsec:be8}. 
Energies are measured from the values at $d_\textrm{max}=8$~fm.

First, let us look into the energies of the $0s$ state; the conventional 
two-$\alpha$ cluster model without the $NN$ correlations (lines ``$0s$'' in Fig.~\ref{fig:ene-g3rs2-v2m-1}). 
The $d$ dependences of the two-$\alpha$ energy 
are similar 
in the G3RS2-3R 
(Fig.~\ref{fig:ene-g3rs2-v2m-1}~(a))
and V2m-3R 
(Fig.~\ref{fig:ene-g3rs2-v2m-1}~(d))
cases; mild repulsive effect
in the long-distance region
caused by the kinetic energy 
(Fig.~\ref{fig:ene-g3rs2-v2m-1}~(b) and (e))
and the attractive effect
in the middle-distance region owing to the central interaction
(Fig.~\ref{fig:ene-g3rs2-v2m-1}~(c) and (f)).
These features are the same as
V2:$0s$ in Fig.~\ref{fig:be8-offset}, but this attractive effect of the central interaction
in G3RS2-3R and V2m-3R
is smaller
than in V2:$0s$ because of the weaker 
triplet-even part of the central interactions. 
As a result, 
the minimum point for the total energy disappears in the conventional $0s$ calculation for G3RS2-3R and V2m-3R.

Next, we compare energies of the frozen-$\alpha$ calculation 
(lines ``frozen'' in Fig.~\ref{fig:ene-g3rs2-v2m-1})
with those for the $0s$ state
(``$0s$'' in Fig.~\ref{fig:ene-g3rs2-v2m-1}). 
The frozen-$\alpha$ calculation shows how the energy of the two-$\alpha$ system changes when 
they just come closer without the internal excitation (without the core polarization).
The energy difference between these two 
shows the reflection
of the high-momentum $NN$ correlations in the ground state of $^4\textrm{He}$
to the two-$\alpha$ energy.
The frozen-$\alpha$ calculation has the repulsive effect
at short distances considerably higher
than $0s$ (Fig.~\ref{fig:ene-g3rs2-v2m-1}~(a) and (d)), which
dominantly comes from the
kinetic term (Fig.~\ref{fig:ene-g3rs2-v2m-1}~(b) and (e)).
Here the repulsive effect of the tensor term is also significantly large
(Fig.~\ref{fig:ene-g3rs2-v2m-2}~(a) and (d)). The energy differences at $d=1$~fm  are
about 100~MeV for the kinetic term and about 30~MeV for the tensor term. 
These repulsive effects  
at short distances ($d\le 2$~fm for the kinetic term and 
$d\le 1$~fm for the tensor term)
are attributed to
the Pauli blocking effect 
for the high-momentum $NN$-correlation 
in the $^3D$ state.
The repulsive effect of the tensor interaction is almost the same extent in G3RS2-3R and V2m-3R, whereas 
that of the kinetic term is slightly higher in G3RS2-3R than in V2m-3R because of  
larger mixing of
high-momentum components in the correlated $S$-wave pair.

Then, we compare the results of the optimized-$\alpha$ calculation 
(lines ``opt'' in Fig.~\ref{fig:ene-g3rs2-v2m-1})
with the 
frozen-$\alpha$ one
(lines ``frozen'' in Fig.~\ref{fig:ene-g3rs2-v2m-1}).
The difference comes from
the core polarization effect, $i.e.$, the internal structure change of the $\alpha$ cluster
caused by the presence of another $\alpha$ cluster.
In the optimized-$\alpha$ calculation, 
the total energy decreases because of the structure change in the $d\lesssim 2$ fm region,
which is indeed seen in the 
significant reduction of the $^3D$ component and enhancement of the $0s$ component 
(Table \ref{tab:be8} and Fig.~\ref{fig:be8-offset}). 
As already shown in the frozen-$\alpha$ calculation, 
the approaching $\alpha$ clusters strongly feel the repulsive effect coming from the kinetic term, but in the optimized-$\alpha$ calculation, 
the structure change allows to
reduce the repulsive effect by suppressing
the high-momentum $NN$-correlations mainly in the $^3D$ channel.
Moreover, in G3RS2, this $D$-state suppression also reduces the attractive effect   
of the central-even (mainly $^3E$) interaction, especially for its middle-range term ($\eta_{c,2}=0.942$ fm), 
as shown in Fig.~\ref{fig:ene-g3rs2-v2m-3}~(b).

The structure change of the $\alpha$ clusters approaching each other
affects not only the internal excitation 
of the clusters but also the relative motion 
between them through the antisymmetrization effect for the nucleons in the two $\alpha$ clusters.
To clearly distinguish these two effects, 
we separate the energy of the optimized-$\alpha$ calculation into the internal 
(lines ``int'')
and external (lines ``ext'') parts
as described previously (in subsection \ref{subsec:internal-external}). 
Their $d$ dependence is shown
in Figs.~\ref{fig:ene-g3rs2-v2m-1}, \ref{fig:ene-g3rs2-v2m-2}, and \ref{fig:ene-g3rs2-v2m-3}.
Note that frozen-$\alpha$ does not contain the internal excitation energy by definition.
Here the reduction of the kinetic energy in optimized-$\alpha$ compared with frozen-$\alpha$ is understood as
the decrease of both the internal kinetic energy and external one
due to the core polarization (Fig.~\ref{fig:ene-g3rs2-v2m-1} (b) and (e));
the reduction of the high-momentum correlations decreases both the internal and external kinetic energies.
On the other hand, for the tensor term, the repulsive effect 
mainly comes from the large reduction 
of the $^3D$ component in the internal part
(Fig.~\ref{fig:ene-g3rs2-v2m-2} (b) and (e)).
For the central interaction, 
the repulsive effect in the long-distance region 
 of the central-odd term 
only weakly
contributes to the external energy 
but not to the internal energy 
(Fig.~\ref{fig:ene-g3rs2-v2m-2} (c) and (f))
because
its contribution to the binding energy of  $^4$He is quite small.
These features are qualitatively 
similar in the G3RS2-3R and V2m-3R results. 

However, for the central-even interaction,
 one can see a clear difference between 
G3RS2-3R and V2m-3R. 
The internal excitation provides significant 
repulsive effect in the region of $d\le 2$~fm 
only in the case of G3RS2-3R (Fig.~\ref{fig:ene-g3rs2-v2m-2}~(b) and (e)).
In G3RS2-3R,
this repulsive effect mainly comes from 
the middle-range  ($\eta_{c,2}=0.942$ fm) term particularly in the $^3E$ channel 
(see Fig.~\ref{fig:ene-g3rs2-v2m-3}~(b)).
As described in the previous section for $^4$He, in G3RS2-3R, 
the tensor correlation, {\it i.e.} the $^3D$ component of the $NN$ correlation,
 contributes  
also to the 
enhancement of the attraction coming from
the middle-range of the central-even interaction 
through the $^3S$-$^3D$ coupling.  
In the $d\le 2$ fm region
of the two-$\alpha$ system,
not only the $^3D$ component but also 
the correlated $S$-wave component is significantly reduced (see Fig.~\ref{fig:be8-offset}).
This means that the $D$-state suppression 
by the Pauli blocking induces the reduction of the $^3S$ component through the $^3S$-$^3D$ coupling
and results in the repulsive effect in the central-even interaction. 
Such an effect cannot be seen in V2m-3R, and this is a characteristic feature of G3RS2-3R 
containing the deep attraction (potential pocket) in the middle-range and the repulsion in the short-range of the central interaction,
which are the origins of stronger repulsive effect
 in the $d\le 2$ fm region
of the $\alpha$-$\alpha$ potential compared with V2m-3R.

\section{Summary} \label{sec:summary}

In the present study, we aimed
to describe the short-range correlation 
caused by the repulsive core of the central interaction in addition to the tensor correlation.
We applied AQCM-T to
$^4\textrm{He}$ and $^8\textrm{Be}$
and in particular, we focused on the
$NN$ correlations in $\alpha$ clusters. 
We used the $NN$ interactions including
realistic ones containing a repulsive core for the central part.
The adopted interactions are
G3RS1-3R and G3RS2-3R, realistic interactions  
with the tensor term and short-range central core, V2m-3R, effective interaction 
with the tensor term but no short-range central core, and V2, effective interaction without the tensor nor short-range central core.

The binding energy and size of ${}^4\textrm{He}$ are reasonably reproduced with G3RS2-3R, V2m-3R, and V2 interactions, but
the energy of each term of the Hamiltonian is different; 
G3RS2-3R and V2m-3R give higher kinetic energy and larger attraction of the potential energy 
than V2 because of the $NN$ correlations. the tensor interaction gives 
significant contribution almost comparable to 
the central interaction. 
 The energy of each term and $D$-state probability of 
${}^4\textrm{He}$ obtained with
G3RS2-3R and V2m-3R are similar,
but some different features can be seen in the $S$-wave $NN$-correlations. 
The $S$-wave pair wave functions of G3RS2-3R show some suppression in the 
short-range ($r\lesssim 2$~fm) region and the enhancement of the peak in the middle-range ($r \sim 1$~fm) region, but those of V2m-3R do not show such features.

We also applied AQCM to the two-$\alpha$ 
cluster structure of ${}^8\textrm{Be}$
and 
compared the results of G3RS2-3R, V2m-3R, and V2, 
while paying attention to the roles of the $NN$ correlations in the two-$\alpha$ potential.
In the cases of G3RS2-3R and V2m-3R, as two $\alpha$ clusters approach each other, 
the attractive contribution of the tensor interaction is drastically reduced because 
the Pauli blocking effect between two $\alpha$ clusters suppresses the $D$-state configuration 
in the $\alpha$ clusters. This tensor suppression gives a repulsive effect 
to the two-$\alpha$ system in the region of distance $d\lesssim 2$~fm.
This can be seen in 
G3RS2-3R and V2m-3R with the tensor terms but not in V2 without
it. 
In G3RS2-3R, another repulsive effect is seen in the same $d$ region,
which comes from the reduction of the 
central-even interaction, but this effect originates from the same $D$-state suppression. 
Namely,  the $D$-state suppression 
induces the decrease of the $^3S$ component through the $^3S$-$^3D$ coupling, 
which results in the reduction of the attractive effect of the central-even interaction
in the middle-range region ($r\sim 1$~fm). 
Such repulsive effect for the contribution of the central-even interaction in $d\lesssim 2$~fm 
cannot be seen in V2m-3R without the repulsive core of the central interaction.

We have shown that the tensor effect can be taken into account by utilizing AQCM-T
not only in $^4$He but also in $^8$Be.
The method can be applied to heavier nuclei 
and useful to
 clarify the interplay between
the tensor correlation and appearance of the cluster structure
in such nuclei as $^{12}$C and $^{16}$O.
Also, further renovation of the model, which is capable of utilizing interactions with higher cores, is ongoing.

\begin{acknowledgments}
The computational calculations of this work were performed by using the
supercomputer at Yukawa Institute for Theoretical Physics, Kyoto University. This work was supported by 
JSPS KAKENHI Grant Numbers  26400270 (Y.~K-E.), 17K05440 (N.~I.), 18J13400 (H.~M.), and 18K03617 (Y.~K-E.).
\end{acknowledgments}


\begin{references}

\bibitem{Brink}
D.~M.~Brink, in {\it Proceedings of the International School of Physics ``Enrico Fermi" Course XXXVI},
edited by C.~Bloch (Academic, New York, 1966), p. 247.

\bibitem{Fujiwara}
Y.~Fujiwara {\it et al.,}
Supple. of Prog. Theor. Phys. {\bf 68}, 29 (1980).

\bibitem{Hoyle}
F.~Hoyle, D.~N.~F.~Dunbar, W.~A.~Wenzel, and W.~Whaling, Phys. Rev. {\bf 92}, 1095c (1953).

\bibitem{Uegaki}
E.~Uegaki, S.~Okabe, Y.~Abe, and H.~Tanaka, Prog. Theor. Phys. {\bf 57}, 1262 (1977).

\bibitem{THSR}
A.~Tohsaki, H.~Horiuchi, P.~Schuck, and G.~R\"{o}pke, Phys.
Rev. Lett. {\bf 87}, 192501 (2001).

%
%
%

\bibitem{ATMS}
M.~Sakai, I.~Shimodaya, Y.~Akaishi, J.~Hiura, and H.~Tanaka,
Supple. of Prog. Theor. Phys. {\bf 56}, 32 (1974)

\bibitem{Kamada}
H.~Kamada {\it et al}., Phys. Rev. C {\bf 64}, 044001 (2001).

\bibitem{TOSM}
T.~Myo, K.~Kat\=o, and K. Ikeda, Prog. Theor. Phys. {\bf 113}, 763 (2005).

%
%
%


\bibitem{Alvioli}
M.~Alvioli, C.~Ciofi~degli~Atti, and H.~Morita,
Phys. Rev. Lett. {\bf 100}, 162503 (2008).

\bibitem{Feldmeier}
H.~Feldmeier, W~Horiuchi, T.~Neff, and Y.~Suzuki,
Phys. Rev. C {\bf 84}, 054003 (2011).

\bibitem{Wiringa}
R.~B.~Wiringa, R.~Schiavilla, S.~C.~Pieper, and J.~Carlson,
Phys. Rev. C {\bf 89}, 024305 (2014).

\bibitem{Atti}
C.~C.~degli~Atti, Phys. Rep. {\bf 590} 1 (2015).

\bibitem{Alvioli-2}
M.~Alvioli, C.~C.~degli~Atti, and H.~Morita,
Phys. Rev. C {\bf 94}, 044309 (2016).

\bibitem{Hen:2016kwk} 
  O.~Hen, G.~A.~Miller, E.~Piasetzky, and L.~B.~Weinstein,
  Rev.\ Mod.\ Phys.\  {\bf 89}, no. 4, 045002 (2017).


\bibitem{KanadaEnyo:1995tb}
  Y.~Kanada-En'yo, H.~Horiuchi, and A.~Ono,
  Phys.\ Rev.\  C {\bf 52}, 628  (1995).

\bibitem{KanadaEnyo:1995ir}
  Y.~Kanada-En'yo and H.~Horiuchi,
  Phys.\ Rev.\  C {\bf 52}, 647 (1995).

\bibitem{AMDsupp} 
Y.~Kanada-En'yo and H.~Horiuchi,
Supple. of Prog. Theor. Phys. {\bf 142},  205 (2001).

\bibitem{KanadaEn'yo:2012bj}
  Y.~Kanada-En'yo, M.~Kimura, and A.~Ono,
 Prog. Theor. Exp. Phys. {\bf 2012},  01A202 (2012).

\bibitem{Dote}
A.~Dot\'e, Y.~Kanada-En'yo,  H.~Horiuchi,  Y.~Akaishi, and K.~Ikeda,
Prog. Theor. Phys. {\bf 115}, 1069 (2006).

\bibitem{Neff}
T.~Neff and H.~Feldmeier, Nucl. Phys. \textbf{A738}, 357 (2004).

\bibitem{Roth}
R.~Roth, T.~Neff, and H.~Feldmeier, Prog. Part. Nucl. Phys. {\bf 65}, 50 (2010).

\bibitem{Chernykh}
M.~Chernykh, H.~Feldmeier, T.~Neff, P. von~Neumann-Cosel, and A.~Richter,
Phys. Rev. Lett.  {\bf 98}, 032501 (2007); {\it ibid} {\bf 105}, 022501 (2010).


\bibitem{Simple}
N.~Itagaki, H.~Masui, M.~Ito, and S.~Aoyama, Phys. Rev. C {\bf 71}, 064307 (2005). 

\bibitem{Itagaki-SMT}
N.~Itagaki, H.~Masui, M.~Ito, S.~Aoyama, and K.~Ikeda,
Phys. Rev. C {\bf  73}, 034310 (2006).

\bibitem{Masui}
H.~Masui and N.~Itagaki, Phys. Rev. C {\bf 75}, 054309 (2007). 

\bibitem{Yoshida2}
T.~Yoshida, N.~Itagaki, and T.~Otsuka, Phys. Rev. C {\bf 79}, 034308 (2009).

\bibitem{Ne-Mg}
N.~Itagaki, J.~Cseh, and M.~P{\l}oszajczak, Phys. Rev. C {\bf 83}, 014302 (2011).

\bibitem{Suhara}
T.~Suhara, N.~Itagaki, J.~Cseh, and M.~P{\l}oszajczak,
Phys. Rev. C {\bf 87}, 054334 (2013).

\bibitem{Suhara2015}
T.~Suhara and Y.~Kanada-En'yo,
Phys. Rev. C {\bf 91}, 024315 (2015).

\bibitem{Itagaki}
N.~Itagaki, H.~Matsuno, and T.~Suhara, Prog. Theor. Exp. Phys. {\bf 2016}, 
093D01 (2016). 

\bibitem{Itagaki-CO}
N.~Itagaki, Phys. Rev. C {\bf  94}, 064324  (2016). 

\bibitem{Matsuno}
H.~Matsuno, N.~Itagaki, T.~Ichikawa, Y.~Yoshida, and Y.~Kanada-En'yo,
Prog. Theor. Exp. Phys. {\bf 2017}, 063D01  (2017). 
 
\bibitem{Matsuno2}
H.~Matsuno and N.~Itagaki,
Prog. Theor. Exp. Phys. {\bf 2017}, 123D05 (2017).

\bibitem{O24}
N.~Itagaki and A.~Tohsaki,
Phys. Rev. C {\bf 97}, 014307 (2018). 






\bibitem{iSMT}
N.~Itagaki and A.~Tohsaki,
Phys. Rev. C {\bf 97}, 014304 (2018). 

\bibitem{hmAMD}
T.~Myo, H.~Toki, K.~Ikeda, H.~Horiuchi, T.~Suhara,
M.~Lyu, M.~Isaka, and T.~Yamada,
Prog. Theor. Exp. Phys. {\bf 2017}, 111D01 (2017).

\bibitem{hmAMD-2}
T.~Myo,
Prog. Theor. Exp. Phys. {\bf 2018}, 031D01 (2018). 


\bibitem{AQCM-T}
H.~Matsuno, Y.~Kanada-En'yo, and N.~Itagaki,
arXiv.1805.10087.

\bibitem{Bando}
H.~Band\=o, S.~Nagata, and Y.~Yamamoto, Prog. Theor. Phys. {\bf 44}, 646 (1970). 

\bibitem{QMC} 
R.~B.~Wiringa, Steven~C.~Pieper, J.~Carlson, and V.~R.~Pandharipande, Phys. Rev. C {\bf 62}, 014001 (2000).

\bibitem{Yamamoto}
Y.~Yamamoto, T.~Togashi, and K.~Kat\=o, Prog. Theor. Phys. {\bf 44}, 646 (2010).

\bibitem{Horii}
K.~Horii, H.~Toki, T.~Myo, and K.~Ikeda,
Prog. Theor. Phys. {\bf 127}, 1019 (2012). 

\bibitem{G3RS} 
R. Tamagaki, Prog. Theor. Phys. {\bf 39}, 91 (1968).

\bibitem{Volkov}
A.~B.~Volkov, Nucl. Phys. \textbf{74}, 33 (1965).

\bibitem{BB}
D. M. Brink and E. Boeker, Nucl. Phys. A {\bf 91}, 1 (1967).



%
%

\bibitem{angeli13}
I.~Angeli and K.~P.~Marinova, At.~Data Nucl.~Data Tables {\bf 99}, 69 (2013).   



\end{references}
\end{document}